\documentclass[12pt]{article}

\usepackage[T1]{fontenc}
\usepackage[utf8]{inputenc}
\usepackage{lmodern}
\usepackage{setspace}
\usepackage{amsmath}
\usepackage{amssymb}
\usepackage{amsfonts}
\usepackage{amsthm}
\newtheorem{assumption}{Assumption}
\usepackage{bm}
\theoremstyle{remark}
\newtheorem{remark}{Remark}[section]
\usepackage{graphicx}
\usepackage{float}
\usepackage{booktabs}
\usepackage{subcaption}
\usepackage{array}

\usepackage{algorithm}
\usepackage{algpseudocode}

\usepackage{tikz}
\usetikzlibrary{arrows, chains, positioning, quotes, shapes.geometric}
\usetikzlibrary{arrows.meta}
\tikzset{
node distance = 8mm and 20mm,
start chain = going below,
arrow/.style = {thick,-stealth},
base/.style = {
    draw, thick,
    minimum width=30mm,
    minimum height=10mm,
    align=center,
    inner sep=1mm,
    outer sep=0mm,
    on chain
},
decision/.style = {diamond, base, aspect=2, inner xsep=0mm},
process/.style = {rectangle, base},
startstop/.style = {rectangle, rounded corners, base},
}

\tikzstyle{line} = [draw, -latex']

\usepackage[numbers]{natbib}

\usepackage[a4paper,margin=1in]{geometry}

\usepackage[colorlinks=true,
            linkcolor=blue,
            citecolor=blue,
            urlcolor=blue]{hyperref}

\title{Causally Interpretable Meta-Mediation Analysis With Missing At Random Mediator and Outcome Data}

\author{
Marie-Félicia Beclin\\
\small EPILOGY, Institut Mondor of Biomedical Research, INSERM U955,\\
\small Université Paris-Est Créteil, France\\
\small \texttt{mariefelicia.beclin@gmail.com}
\and
Apolline Courrèges-Vartanian\\
\small Mines Paris, Université Paris Sciences Lettres, France\\
\small \texttt{apolline.courreges-vartanian@etu.mineparistech.fr}
\and
Geneviève Lefebvre\\
\small Department of Mathematics, Université du Québec à Montréal, Canada\\
\small \texttt{lefebvre.gen@uqam.ca}
\and
Tat-Thang Vo\\
\small EPILOGY, Institut Mondor of Biomedical Research, INSERM U955,\\
\small Université Paris-Est Créteil, France\\
\small \texttt{tat-thang.vo@u-pec.fr}
}

\begin{document}

\maketitle
\newpage
\begin{abstract}
 Meta-analyzing natural indirect effect estimates from multiple studies is increasingly used to synthesize evidence on causal pathways of interest. However, standard mediation meta-analysis approaches are typically based on structural equation modeling, which fails to account for mediator-outcome confounding, is not readily extended to address missing mediator and outcome data, and is often unclear about the target population to which the summary indirect effect pertains. In this work, we propose a novel method that addresses these limitations. Our approach transports study-specific natural indirect effect estimates to a well-defined target population prior to evidence synthesis. The proposed methods enable the integration of studies that do not explicitly investigate mediation but collect data on the mediator to improve extensiveness. Using semiparametric theory, we construct flexible, data-adaptive estimators for the target parameter. Novel random-effects models and non-parametric analogues based on ANOVA sums of squares are also developed to decompose between-study heterogeneity into distinct sources that may affect the causal interpretability of the obtained findings. Finite-sample performance of the proposed methods is evaluated through simulated and real-world data. 
\end{abstract}

\bigskip

\noindent
\textbf{Keywords:}
Mediation Analysis, Natural Indirect Effect, Meta-analysis, Semiparametric theory.

\section{Introduction}

Mediation analysis is a widely used statistical framework for assessing the mechanisms through which an exposure or treatment affects an outcome. The analysis aims to decompose the total effect of the exposure into effects operating through a candidate mediator and effects operating independently of it. Foundational work by \citet{pearl2011direct} and \citet{robins1992identifiability} introduced the concepts of natural direct and indirect effects, which are identifiable under single- and cross-world independence assumptions encoded within a nonparametric structural equation model with independent errors (NPSEM-IE) \citep{pearl2009causality}. Estimation strategies such as the G-formula and propensity score weighting have been widely applied \citep{rosenbaum1983central, imbens2015causal, hirano2003efficient}, while more recent developments in semiparametric theory and influence function–based methods have led to efficient estimators that offer increased robustness to model misspecification \citep{tchetgen2011causal, tchetgen2012semiparametric}.

The strong methodological foundation of mediation analysis has led to a rapid increase in mediation studies across a wide range of scientific disciplines \citep{mackinnon2007mediation, preacher2015advances, richiardi2013mediation, vanderweele2016mediation}. Consequently, systematic reviews and meta-analyses of mediation studies are increasingly being conducted to synthesize evidence regarding the role of specific mediators \citep{vo2020conduct, lee2015does, deneault2023does}. However, the methodology for meta-mediation analysis (MMA) remains underdeveloped, leaving substantial room for methodological advancement. Existing approaches to MMA primarily rely on meta-analytic structural equation modeling (MASEM) \citep{cheung2016random} or parametric marginal likelihood (ML) methods \citep{huang2016statistical}. Originating largely from the psychology and social science literature, these approaches often fail to account for mediator–outcome confounding and are generally not adaptable to settings involving nonlinear relationships, such as treatment–mediator interactions or non-continuous mediators and outcomes \citep{vanderweele2016mediation,jiang2015difference}.  
In addition, as extensions of standard random-effect meta-analysis models, MASEM and ML methods generally do not support causal interpretation, as they remain implicit about the target population to which the summary indirect effect pertains. This concern is important when mediator and outcome effect modifiers are differentially distributed across studies \citep{dahabreh2023efficient, vo2019novel, vo2025integration}. 

Recent advances in evidence synthesis have emphasized the importance of transporting causal effect estimates to a well-defined target population prior to aggregation \citep{dahabreh2023efficient, vo2019novel, vo2025integration, rott2024causally, hong2025estimating}. However, this causally interpretable meta-analysis framework has not yet been formally developed for MMA. This setting that is inherently more challenging due to the need to integrate information from multiple components of the mediation pathway. For instance, it is common that primary studies may only investigate the treatment–mediator or mediator–outcome association \citep{vo2022challenges, vo2020conduct, murillo2022psychologically}. Although such studies do not directly estimate the indirect effect, they may still provide informative evidence regarding the plausibility and magnitude of the underlying pathway. Incorporating these studies into a causal MMA framework could substantially expand the evidence base and improve the informativeness of synthesized findings, but doing so requires formal methodological development \citep{vo2022challenges}.

In this paper, we address the above challenges by proposing novel methods for MMA that enable the evaluation of indirect exposure effects for a well-defined target population. Specifically, we develop a two-stage framework in which the target indirect effect is estimated by integrating information on the exposure–mediator and mediator–outcome relationships obtained from different eligible studies. To avoid restrictive parametric assumptions, we derive the efficient influence function for the target-specific indirect effect and leverage semiparametric efficiency theory to construct flexible estimators that can incorporate modern data-adaptive approaches. We further propose meta-analytic approaches for synthesizing standardized indirect effects, which enable the quantification of two distinct sources of between-study variability: outcome-related heterogeneity and mediator-related heterogeneity. Finally, we apply the proposed methods to examine the role of health in the causal relationship between higher education and self-reported life satisfaction worldwide, while investigating heterogeneities across different countries using data from the 2017--2021 World Values Survey, Wave 7, Master Survey Questionnaire \citep{haerpfer2020world}.

\section{Transportability of natural direct effect and indirect effects}\label{section1}
Consider \(S=1,\ldots,K\) mediation studies that evaluate the role of an intermediate variable \(M\) in explaining the causal effect of a binary exposure \(A\) on an outcome \(Y\). In addition, consider \(S=K+1,\ldots,P\) studies that evaluate the causal effect of \(A\) on the mediator \(M\). In this work, we focus on a single mediator and outcome; however, no assumptions are imposed regarding the nature of the mediator, which may be binary, discrete, or continuous (see below). Let \(\boldsymbol{C}\) denote a set of baseline covariates that are commonly measured across studies. Within each study, we allow for missing mediator and outcome data. Specifically, let $R$ and $\tilde{R}$ denote the missingness indicator for $M$ and $Y$, respectively. Here, $R$ (or $\tilde{R}$) takes value $0$ if $M$ (or $Y$) is observed and $1$ otherwise. To account for missing data, we impose the following assumption on the missingness mechanism within each study:
\begin{assumption}\label{ass:mar}(Missing At Random)\label{mar_assumption} $R   \perp (M, Y) \mid A, \boldsymbol{C}, S$ and $\tilde{R} \perp Y      \mid M, A, \boldsymbol{C}, R, S.$
\end{assumption}

\subsection{Pre-existing methods} Three main approaches have been proposed to meta-analyze mediation findings. The first approach, correlation-based MASEM, pools study-specific correlations. For each study i, the observed correlation vector is defined as $\mathbf{r}_i = (\widehat{r}_{AY}, \widehat{r}_{AM}, \widehat{r}_{MY})^\top$
and a random-effects model is used  $
\mathbf{r}_i = \boldsymbol{\rho} + \mathbf{u}_i + \mathbf{\epsilon}_i,$
with $\boldsymbol{\rho}$ the average correlation vector, $\mathbf{u}_i$ a normal random effect reflecting between-study heterogeneity and $\mathbf{\epsilon}_i$ the residual error. Once the correlation model is fitted,
the estimated average correlation matrix $\hat{\bm \rho}$ and the corresponding (estimated) covariance matrix $\bm V$ are used to fit the mediation model by weighted least squares \citep{cheung2014fixed}. Notably, this approach can accommodate studies with partially missing correlation coefficients. For example, studies reporting only the exposure–mediator association can still be incorporated by assuming that the treatment/mediator-outcome correlations in these studies are missing at random. However, confounding adjustment is not naturally accommodated within this framework. In practice, implementations often implicitly assume the absence of mediator-outcome confounding beyond that induced by treatment itself. In addition, handling individual-level missing data, as considered in our setting, is generally difficult.

The second approach, based on marginal likelihood methods, models the study-specific mediator and outcome regressions as $\mathbb{E}(M \mid A, S=i) = \alpha_{0i}+\alpha_iA$ and
$\mathbb{E}(Y \mid A, M, S=i) = \beta_{0i}+\beta_{1i}A+\beta_iM,$
with random effects  
$(\alpha_i,\beta_i) \sim \mathcal{N}\bigl((\mu_\alpha,\mu_\beta),\Sigma\bigr).$
The pooled indirect effect is then estimated using the product-of-coefficients \citep{baron1986moderator} approach, i.e.,  
$\widehat{\theta} = \widehat{\mu}_\alpha \widehat{\mu}_\beta.$ Although straightforward to implement in linear settings, extending this framework to nonlinear models is considerably more challenging, since the natural indirect effect within each study is no longer a simple product of regression coeffcients. Accounting for missing data within each study is also challenging.

The third approach, parameter-based MASEM, applies a standard random-effects model directly to the study-specific indirect effect estimates $\hat\theta_i$, i.e., $\widehat{\theta}_i = \theta + u_i + \epsilon_i,$ where $\theta$ denotes the pooled indirect effect, $u_i$ denotes a study-specific random effect capturing between-study heterogeneity, and $\epsilon_i$ denotes sampling error \citep{cheung2016random}. This approach requires confounding and missing data to be appropriately addressed within each eligible study to ensure valid estimation of the summary indirect effect $\theta$. A key limitation, however, is that it can only incorporate studies reporting a formal mediation analysis together with an associated indirect effect estimate. In practice, this restriction may introduce selective reporting bias, as mediation analyses are often conducted as secondary analyses only when the total effect of $A$ on $Y$, or the component associations between $A$ and $M$ and between $M$ and $Y$, are statistically significant. Consequently, excluding studies that report only partial information may lead to distorted or overly optimistic conclusions.

While the above approaches have been proposed primarily for aggregate data MMAs (AD-MMAs), methods for individual participant data MMAs (IPD-MMAs) have also been developed \citep{saunders2023past, zhang2015developing, hiensch2021stepping, huh2022structural}. For instance, one-stage IPD-MMAs typically rely on parametric models for the mediator and outcome, which incorporate random effects on model coefficients to account for clustering across studies. The summary indirect effect is then commonly estimated using a product-of-coefficients approach \citep{baron1986moderator}, which is difficult to extend to non-linear settings.

A common limitation of all MMA approaches described above, regardless of whether they use aggregate or individual participant data, is that they derive summary indirect effect estimates as weighted averages across studies. Consequently, the causal interpretation of these pooled estimates may become unclear when included studies involve populations with heterogeneous case-mix. Moreover, existing approaches provide only a limited characterization of heterogeneity, as they do not permit a comprehensive decomposition of the distinct sources of between-study heterogeneity. In particular, they do not distinguish whether variability in indirect effect estimates is primarily driven by heterogeneity in the treatment–mediator pathway (\(A \rightarrow M\)) or in the mediator–outcome pathway (\(M \rightarrow Y\)). To overcome these limitations, in the following section, we instead consider transporting study-specific indirect effect information to a well-defined target population before synthesizing the resulting transported effects. This enables a clearer population-level causal interpretation of the summary indirect effect estimate, alongside a more insightful assessment of clinical and methodological heterogeneity across studies.

\subsection{Introducing novel estimands}
In meta-analysis practice, it is widely recognized that, beyond differences in case-mix, eligible studies may also vary with respect to the treatment versions being evaluated \citep{vo2019novel, vo2020conduct}. Such heterogeneity may arise, for example, when pharmacological interventions are administered at slightly different doses or through different routes of administration, or when behavioral interventions differ in their implementation protocols. Explicitly acknowledging and quantifying this source of heterogeneity is generally more informative than ignoring it and naively pooling individual-level data across studies.

To account for treatment version variation, let $Y(a,k,m)$ denote the potential outcome if a patient receives version $k$ of treatment $a$ (evaluated in study $S=k$) and the mediator were set to value \(m\). Likewise, let $M(a,p)$ denote the potential value of the mediator under version $p$ of treatment \(a\).  The nested counterfactual of interest is $Y\bigl(a, k, M(a^{*}, p)\bigr)$, which describes the potential outcome value when a patient receives version $k$ of treatment $a$, while experiencing the mediator value (s)he would have had under version $p$ of treatment $a^*$. Letting $\theta_{k,p}^{a,a^*} =  \mathbb{E}\Bigl[Y\bigl(a, k, M(a^{*}, p)\bigr) \mid S=0\Bigr]$, our focus is on evaluating the estimand:  
$$\zeta_{k,p} = \theta_{k,p}^{a,a}- \theta_{k,p}^{a,a^*},$$
which describes the expected change in the outcome in an external target population $S=0$, when patients in this population are given treatment version $(a,k)$, but the mediator is changed from values naturally observed under treatment version $(a,p)$ to treatment version $(a^*,p)$ with  $a^*\ne a$. 

\begin{remark}\label{remark:versiontreatment}
    When there is no treatment version heterogeneity, in the sense that $Y(a,k,m)=Y(a,m)$ and $M(a,p)=M(a)$ for all $k,p=1,\ldots,K$, $Y\bigl(a, k, M(a^{*}, p)\bigr)$ reduces to the standard counterfactual $Y(a,M(a^*))$ and $\zeta_{k,p}$ reduces to the conventional natural indirect effect in the target population $S=0$.
\end{remark}

\subsection{Causal assumptions and identifiability}
To identify $\theta_{k,p}^{a,a^*} $ from the observed data, we consider the following set of causal assumptions:
\begin{assumption}[Ignorability]
\label{ass:unconfoundness} 
$Y(a,k,m)\ \perp\ A \mid \boldsymbol{C},S=k$ and $M(a,p)\ \perp\ A \mid \boldsymbol{C},S=p.$
\end{assumption}
\begin{assumption}[Consistency]
\label{ass:consistency} 
$Y = Y(a,k,m)$ if $(A,S,M)=(a,k,m)$ and $M = M(a,p)$ if $(A,S)=(a,p).$
\end{assumption}
\begin{assumption}[Cross-world independence]\label{ass:crossworld} $Y(a,k,m)\ \perp\ M(a^*,p)\mid \boldsymbol{C}.$
\end{assumption}

\begin{assumption}[Transportability] \label{ass:transportability} 
$Y(a,k,m)\ \perp\ S \mid \boldsymbol{C}$ and $M(a,p)\ \perp\ S \mid \boldsymbol{C}.$
\end{assumption}

\begin{assumption}[Positivity]\label{ass:positivity}
$0<\mathbb{P}(A=a\mid \boldsymbol{C},S), \mathbb{P}(M=m\mid A,\boldsymbol{C},S),\mathbb{P}(S=s\mid \boldsymbol{C})<1,$ for all relevant $(a,m,s)$. 
\end{assumption}

These assumptions are satisfied when the relationships among the variables follow the causal diagram depicted in Figure 1, which encodes an NPSEM-IE. Intuitively, Assumption \eqref{ass:unconfoundness} requires the absence of unmeasured confounding of both the treatment–mediator and mediator–outcome relationships, while Assumption \eqref{ass:crossworld} rules out mediator–outcome confounders that are themselves affected by treatment. 
To enable the transport of study $k$-specific information to a common target population, Assumption \eqref{ass:transportability} further requires that all effect modifiers of the relationships between treatment version $k$ and the mediator and outcome that are differentially distributed between study population $S=k$ and the target population $S=0$ are measured in $\boldsymbol{C}$.

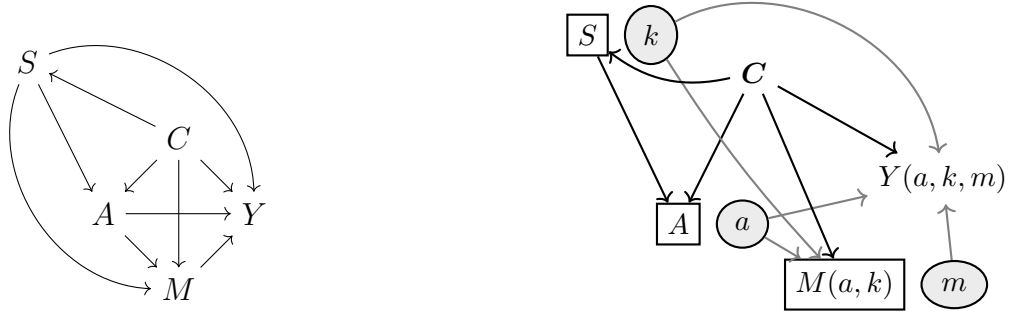
\begin{figure}[ht]
\begin{subfigure}[t]{0.45\textwidth}
      \centering
      \begin{tikzpicture}[node distance =2 cm and 2 cm]
        \node[] (a) at (1,1) {$A$};
        \node[] (y) at (3,1) {$Y$};
        \node[] (m) at (2,0) {$M$};
        \node[] (c) at (2,2) {$C$};
        \node[] (s) at (0,3) {$S$};
        \draw (a) edge[->] (y);
        \draw (c) edge[->] (a);
        \draw (c) edge[->] (m);
        \draw (c) edge[->] (y);
        \draw (a) edge[->] (m);
        \draw (m) edge[->] (y);
       \draw[->, bend left=55] (s) to (y);
        \draw[->] (s) to (a);
        \draw[->, bend right=55] (s) to (m);
        \draw[->] (c) to (s);
      \end{tikzpicture}
\end{subfigure}
\hfill
\begin{subfigure}[t]{0.45\textwidth}
\centering
\begin{tikzpicture}[
    node distance = 2cm and 2cm,
    every node/.style = {font=\small, align=center},
    var/.style = {rectangle, draw, thick, minimum size=4mm},
    fixed/.style = {ellipse, draw, thick, minimum size=5mm, fill=gray!15},
    arrow/.style = {-{Latex[length=3mm,width=2mm]}, thick}
]

\node[var] (A) at (0,0) {$A$};
\node[fixed, right=0.2cm of A] (a) {$a$};

\node[] (C) at (1,2) {$\boldsymbol{C}$};

\node[var] (S) at (-1.2,2.5) {$S$};
\node[fixed, right=0.2cm of S] (s){$k$};

\node[var] (M) at (2.2,-0.8) {$M(a,k)$};
\node[fixed, right=0.2cm of M] (m){$m$};

\node[] (Y) at (3.5,0.6) {$Y(a,k,m)$};

\draw[arrow, draw=gray] (a)  edge[->] (M);
\draw[arrow, draw=gray] (a)  edge[->] (Y);
\draw[arrow, draw=gray] (m)  edge[->] (Y);
\draw[arrow] (C) edge[->] (A);
\draw[arrow] (C) edge[->] (M);
\draw[arrow] (C) edge[->] (Y);
\draw[arrow, bend left=55, draw=gray] (s)  edge[->] (Y);
\draw[arrow] (S) edge[->] (A);
\draw[arrow, bend right=5, gray] (s) edge[->] (M);
\draw[arrow,  bend left=25] (C) edge[->](S);
\end{tikzpicture}
\end{subfigure}
   \caption{Causal diagram and the associated Single World Intervention Graph when intervening on the treatment, mediator and study indicator. $A$: treatment, $Y$: outcome, $\boldsymbol{C}$: baseline confounders, $M$: mediator, $S$: study indicator. The arrow from $S$ to $A$ reflects differences in treatment assignment mechanisms across studies. Arrows from $S$ to $Y$ and $M$ capture study-specific methodological differences affecting the outcome and mediator. 
}
   \label{dag_and_swig}
\end{figure}
Under the proposed causal assumption, $\theta_{k,p}^{a,a^{*}}$ can be identified as:
    \begin{equation} \label{G-formula}
        \theta_{k,p}^{a,a^{*}} =\mathbb{E}\Biggl [\sum_{m \in \mathcal{M}}  \mathbb{E}  (Y \mid \boldsymbol{C}, m, a,S=k)\cdot \mathbb{P}(m \mid \boldsymbol{C},a^{*}, S=p)   \bigg | S=0 \Biggr]
    \end{equation}
    This representation shows that $\theta_{k,p}^{a,a^{*}}$ can be estimated by combining outcome information from study $S=k$ with mediator information from study $S=p$, while standardizing both quantities to the covariate distribution of the target population $S=0$. 
    
Some remarks are noteworthy here. First, in the presence of missing mediator and outcome data, identification of $\theta_{k,p}^{a,a^{*}}$ can still be achieved using only the observed mediator and outcome values, provided that the missingness mechanism within each study satisfies the MAR assumption \eqref{ass:mar} stated previously, i.e.:
    \begin{equation} \label{G-formula1}
        \theta_{k,p}^{a,a^{*}} =\mathbb{E}\Biggl [\sum_{m \in \mathcal{M}}  \mathbb{E}  (Y \mid \boldsymbol{C}, m, a,\tilde{R}=0,S=k)\cdot \mathbb{P}(m \mid \boldsymbol{C},a^{*}, R=0,S=p)   \bigg | S=0 \Biggr].
    \end{equation}
Second, in the absence of treatment-version heterogeneity (Remark \ref{remark:versiontreatment}), Assumptions \eqref{ass:unconfoundness}–\eqref{ass:crossworld} reduce to the standard identification assumptions for the conventional natural indirect effect in the target population. Likewise, Assumption \eqref{ass:transportability} simplifies to the more familiar transportability condition, i.e. $\{Y(a,m), M(a)\}\perp S\mid \boldsymbol{C}$, which implies that the mediator and outcome generating mechanisms are homogeneous across populations. In particular, $\mathbb{E}(Y \mid \boldsymbol{C}, M, A,S=k)= \mathbb{E}(Y \mid \boldsymbol{C}, M, A,S=k')$ and $\mathbb{P}(m \mid \boldsymbol{C},a^{*}, S=p)=\mathbb{P}(m \mid \boldsymbol{C},a^{*}, S=p')$ for $k\ne k'$ and $p\ne p'$. Under such homogeneity, the target expectation is identifiable via:
    \begin{equation*}
        \mathbb{E}\{Y(a,M(a^*))\mid S=0\} = \mathbb{E}\Biggl [\sum_{m \in \mathcal{M}}  \mathbb{E}  (Y \mid \boldsymbol{C}, m, a,S>0) \cdot \mathbb{P}(m \mid \boldsymbol{C},a^{*}, S>0)   \bigg | S=0 \Biggr].
    \end{equation*}
Thus, rather than transporting mediator and outcome information separately from each study to the target population, one may instead fuse data across studies and transport the combined information to $S=0$. However, such an approach has two important limitations. First, when mediator or outcome data are missing, pooled analyses implicitly require the missingness mechanisms to be homogeneous across studies, namely $R\perp M|A,\boldsymbol{C},S>0$ and $\tilde{R}\perp Y|A, M, \boldsymbol{C},S>0$, which may be unrealistic in practice. By contrast, separately transporting mediator and outcome information from each study allows missing data mechanisms to be handled study-by-study prior to transport. More importantly, the proposed framework enables explicit assessment of mediator- and outcome-related heterogeneity across populations. For example, testing whether $\theta_{1,p}^{a,a^{*}}=\ldots=\theta_{K,p}^{a,a^{*}}$ provides an indirect assessment of whether the outcome model $\mathbb{E}(Y|A, M, \boldsymbol{C}, S=k)$ is homogeneous across studies, because the compared quantities share the same mediator component $\mathbb{P}(m\mid A, \boldsymbol{C}, S=p)$. Analogous comparisons can be used to evaluate heterogeneity in the mediator component. Measures quantifying the magnitude of these distinct forms of heterogeneity can also be developed, allowing investigators to assess whether important between-study differences exist in either the mediator or outcome mechanisms underlying the indirect effect (see below). Such evaluations are generally not possible when data are simply fused.
\subsection{Parametric estimation strategies}
We now discuss different estimation strategies for $\theta_{k,p}^{a,a^*}$, based on the identification formula \eqref{G-formula1}. The available data include $n_k$ observations from each source population $k = 1\ldots K$, of the form $\left(A_i, Y_i,M_i, \boldsymbol{C}_i, S_i=k, R_i, \tilde{R}_i, \right)$,  $n_p$ observations from each source population $k = K+1\ldots P$, of the form $\left(A_i, M_i, \boldsymbol{C}_i, S_i=p, R_i \right)$ and $n_0$ observations from the target population, for which only $\left(\boldsymbol{C}_i, S_i=0\right)$ are observed. The total sample size is $n=\sum_{k=0}^{P} n_k$, where $k=0$ denotes the target population. To reduce notational burden, we denote $I(k)$ for $I(S=k)$ and so forth. For $a,a^*=0,1$, $s'=0,\ldots, K$ and $s,k,p=1,\ldots, K$, we define $\bm \eta =(U_s, V_s, \tau^{a}_{s}, \pi^{a}_{s},  \tilde{\rho}^{a}_{s}, \rho^{a}_{s}, Q^{a}_{s}, b^{a, a^*}_{k,p})$ the vector of nuisance parameters, where $U_{s'}=U_{s'}(\boldsymbol{C})=\mathbb{P}(S=s' \mid C)$; $V_s=V_s(\boldsymbol{C}, M)=\mathbb{P}(S=s \mid \boldsymbol{C}, M, R=0)$; $\pi^{a}_{s}=\pi^{a}_{s}(\boldsymbol{C})=\mathbb{P}(A=a \mid \boldsymbol{C},s)$; $\tau^{a}_{s}=\tau^{a}_{s}(\boldsymbol{C},M)= \mathbb{P}(A=a \mid \boldsymbol{C}, M, R=0,s)$; $\tilde{\rho}^{a}_{s}=\tilde{\rho}^{a}_{s}(M,\boldsymbol{C}) = \mathbb{P}(\tilde{R}=0 \mid a, M, \boldsymbol{C}, R=0, s)$; $\rho^{a}_{s}=\rho^{a}_{s}(\boldsymbol{C}) = \mathbb{P}(R=0 \mid a,C,s)$; $Q_{s}^{a}=Q_{s}^{a}(\boldsymbol{C}, \boldsymbol{M})=\mathbb{E}(Y \mid a, \boldsymbol{C}, M,s)$  and $b^{a, a^*}_{k,p}=b^{a, a^*}_{k,p}(\boldsymbol{C})=\mathbb{E}\{Q_{k}^{a}(\boldsymbol{C},M) \mid a^*, C,p\}$. 

In the so-called G-compuation approach, we first postulate and fit a parametric outcome model for 
$Q_{k}^{a}(\boldsymbol{C},M)$ on data of individuals receiving treatment $a$ in study $k$. The fitted model is then used to predict $\widehat{Q}_{k}^{a}(\boldsymbol{C}_i, M_i)$ for all individuals $i$ in population $p$ with treatment $a^*$ who have observed mediator values.  A second parametric model is then postulated for $b_{k,p}^{a, a^*}(\boldsymbol{C})$, which can be fitted using $\widehat{Q}^{a}_{k,p}(\boldsymbol{C}_i, M_i)$ and data on $C$ of subjects treated by $a^*$ in population $p$.
The resulting G-formula estimator for $\theta_{k,p}^{a,a^*}$, obtained by applying this model to the target population $S=0$, is
$\widehat{\theta}^{\mathrm{Gformula}}_{a,a^*;n}
=
\frac{1}{n_0}\sum_{i:S_i=0}
\widehat{b}^{a,a^*}_{k,p}(\boldsymbol{C}_i).$ 

Alternatively, $\theta_{k,p}^{a,a^*}$ can also be estimated by inverse weighting upon noting that:
\begin{equation}\label{IPW_1}
   \theta^{a,a^{*}}_{k,p}(P)  =   \frac{1}{P_0}\;
   \mathbb{E}\Biggl[ \Omega^{a,a^{*}}_{k,p}(\boldsymbol{C},M) \, \, 
       Y\, I(a,k,\tilde{R}=0) \Biggr].
\end{equation}
where $P_0 = \mathbb{P}(S=0)$. An equivalent representation is given by
\begin{equation}\label{IPW_2}
    \theta^{a,a^{*}}_{k,p}(P)  =  \frac{1}{P_0} \mathbb{E}\biggl[   W^{a^{*}}_{p}(\boldsymbol{C}) \, \,I(a^*,p,R=0) \, \,  \mathbb{E} \left[Y \mid \boldsymbol{C}, M, a, S=k, \tilde{R}=0  \right ]  \biggr]
\end{equation}
where
\begin{eqnarray*}
    \displaystyle \Omega^{a,a^{*}}_{k,p} (\boldsymbol{C},M) &=&\frac{ \tau^{a^*}_{p}(\boldsymbol{C},M) \,\,   V_p(\boldsymbol{C},M) \, \, U_0(\boldsymbol{C})  }{\tau^{a}_{k}(\boldsymbol{C},M) \,\,   V_k(\boldsymbol{C},M) \, \pi_{a^*,p}(\boldsymbol{C}) \, U_p(\boldsymbol{C}) \, \,  \tilde{\rho}^{a}_{k}(M,\boldsymbol{C}) \,\,   \rho^{a^*}_{p}(\boldsymbol{C}) } \\
\displaystyle W^{a^{*}}_{p}(\boldsymbol{C})&=& \frac{U_0(\boldsymbol{C} ) }{U_p(\boldsymbol{C}) \,  \pi^{a^*}_{p}(\boldsymbol{C}) \,\,   \rho^{a^*}_{p}(\boldsymbol{C})} 
\end{eqnarray*}
Proof of Equations \eqref{IPW_1} and \eqref{IPW_2} are available in the Supplementary Materials. The inverse weighting representation in Equation \eqref{IPW_1} requires parametric specification of $U_0$, $U_p$, $V_p$, $V_k$, $\tau^{a}_{k}$, $\tau^{a^*}_{p}$, $\pi^{a}_{k}$, $\pi^{a^*}_{p}$, and the missingness mechanisms $\tilde{\rho}^{a}_{k}, \rho^{a^*}_{p}$. By contrast, the representation in Equation~\eqref{IPW_2} relies on specification of $U_0,U_p,\pi^{a^*}_{p}, \pi^{a^*}_{p}$ and the missingness mechanisms $\rho^{a^*}_{p}.$    Consequently, the two representations motivate distinct estimation strategies with different robustness and modeling requirements. Standard errors of the resulting estimators for $\theta_{k,p}^{a,a^*}$ can be obtained by using M-estimation theory \citep{huber1992robust}. 


\subsection{Data-adaptive estimation strategies}
Parametric estimation approaches suffer from an important limitation: they generally fail to provide consistent estimators of $\theta_{k,p}^{a,a^*}$ when the underlying parametric models are misspecified. In this section, we therefore leverage semiparametric theory to develop efficient, data-driven estimation procedures for $\theta_{k,p}^{a,a^*}$. These approaches can achieve $\sqrt{n}$-rate convergence (where $n$ denotes the total sample size across the populations) to the target parameter even when nuisance functions are estimated at slower rates, for example through flexible machine learning methods.


To construct such estimators, we first characterize the efficient influence function (EIF), denoted $\varphi_{k,p}^{a,a^*}$, of the target parameter $\theta_{k,p}^{a,a^*}$. For a fixed nuisance parameter value $\bm  \eta'$ (corresponding, for instanceto a preliminary estimator or its probability limit),  for $\theta' = \theta_{k,p}^{a,a^*}(\bm  \eta')$ and $\theta_0 = \theta_{k,p}^{a,a^*}(\bm  \eta)$  the EIF yields the expansion
$$
\theta'-\theta_0
=
-\mathbb{E}\{\varphi_{k,p}^{a,a^*}(O;\bm \eta', \theta')\}
+
R(\bm  \eta, \bm \eta'),
$$

where $R(\bm  \eta,\bm  \eta')$ is a second-order remainder term that can typically be expressed as sums of products of estimation errors of the form
$\mathbb{E}\Bigl[
c(\bm \eta,\bm \eta')
\{f(\bm \eta')-f(\bm \eta)\}
\{g(\bm \eta')-g(\bm \eta)\}
\Bigr].$ The first-order terms $-\mathbb{E}\{\varphi_{k,p}^{a,a^*}(O; \bm \eta', \theta'\}$ can be rewritten as $ \mathbb{P}_n \{ (\varphi_{k,p}^{a,a^*}(O;\bm \eta, \theta_0) - \varphi_{k,p}^{a,a^*}(O;\bm \eta', \theta')) \} +   (\mathbb{P}_n -P) \{ \varphi_{k,p}^{a,a^*}(O;\bm \eta', \theta') - \varphi_{k,p}^{a,a^*}(O;\bm \eta, \theta_0) \}.$ Under certain conditions (discussed later) on the estimated nuisance parameters, the first-order bias of the G-formula estimator $\theta_{k,p}^{a,a^*}(\bm \eta')$ is reduced as $\mathbb{P}_n\{\varphi_{k,p}^{a,a^*}(O;\bm \eta', \theta')\},$ typically called the G-formula bias. 
The EIF therefore naturally motivates the construction of debiased estimators such as the One-Step (OS) estimator and Targeted Maximum Likelihood Estimator (TMLE). When the remainder term $R(\bm \eta,\bm \eta')$ is sufficiently small, these estimators can be shown to achieve desirable asymptotic properties, including consistency, asymptotic normality, and semiparametric efficiency.


In the Online Supplementary Materials, we prove that the EIF of $\theta_{k,p}^{a,a^*}$ can be expressed as:
    \begin{align} \label{eif}
     \varphi^{a,a^*}_{k,p}&=  \frac{\Omega_{k,p}^{a,a^*}(\boldsymbol{C}, \boldsymbol{M}) I(a,k,\tilde{R}=0)}{P_0}  \bigl( Y - Q_{k}^{a}(\boldsymbol{C}, \boldsymbol{M})  \bigr)\\ & +  \frac{W_{p}^{a^*}(\boldsymbol{C}) I(a^*,p, R=0)}{P_0}  \bigl( Q_{k}^{a}(\boldsymbol{C}, M) - b_{k,p}^{a,a^*}(\boldsymbol{C}) \bigr) 
     + \frac{I(0)}{P_0}  \bigl( b_{k,p}^{a,a^*}(\boldsymbol{C}) -   \theta_{k,p}^{a,a^*}  \bigr) \notag
\end{align}
The EIF of $\zeta_{k,p}$ is $\varphi^{a,a}_{k,p}-\varphi^{a,a^*}_{k,p}$, as a result of the delta method for influence functions (see, e.g., \cite{kennedy2024semiparametric}, Section~3.4.3). 
While our primary focus is on the indirect effect defined on the difference scale, alternative definitions based on other effect scales can also be considered, such as the risk ratio or odds ratio when the outcome is binary. Another causal mediation measure that is commonly used in practice is the proportion of the treatment effect mediated by \(M\), defined as 
$PTE = {NIE}/{TE}$,
which is meaningful when the total effect is non-null \citep{tsiatis1995modeling,wang2002measure,lecoent2025}. Using the delta-method, we derive the EIFs corresponding to indirect effects defined on these alternative scales (Table~\ref{table:eif}). Below, we construct OS and TMLE estimators for the indirect effect defined on the difference scale. Analogous estimators for alternative scales can be developed similarly, and details are therefore omitted.

\begin{table}[ht]

\caption{Natural indirect effect (NIE) and proportion of treatment effect (PTE) estimands with their EIF. OR-PTE simplifies to the natural direct effect on the OR scale.}
\label{table:eif}
\centering
\small
\renewcommand{\arraystretch}{1.8}

\begin{tabular}{|c|m{3.2cm}|m{8.5cm}|}
\hline
\textbf{Causal Effect} & \centering \textbf{Link function $h$} & \centering \textbf{EIF} \tabularnewline
\hline

RD-NIE$(a)$ 
& $\theta_{a,1}-\theta_{a,0}$ 
& $\varphi_{a,1}-\varphi_{a,0}$ \\
\hline

RR-NIE$(a)$ 
& $\dfrac{\theta_{a,1}}{\theta_{a,0}}$
& $\dfrac{1}{\theta_{a,0}}
\left(
\varphi_{a,1}
-
\dfrac{\theta_{a,1}}{\theta_{a,0}}\varphi_{a,0}
\right)$ \\
\hline

OR-NIE$(a)$ 
& $\dfrac{\theta_{a,1}}{1-\theta_{a,1}}
\dfrac{1-\theta_{a,0}}{\theta_{a,0}}$
& $\dfrac{1}{\theta_{a,0}(1-\theta_{a,1})}
\left(
\dfrac{1-\theta_{a,0}}{1-\theta_{a,1}}\varphi_{a,1}
-
\dfrac{\theta_{a,1}}{\theta_{a,0}}\varphi_{a,0}
\right)$ \\
\hline

RD-PTE$(a)$ 
& $\dfrac{\theta_{a,1}-\theta_{a,0}}
{\theta_{1,1}-\theta_{0,0}}$
& $\dfrac{1}{\theta_{1,1}-\theta_{0,0}}
\left(
\varphi_{a,1}-\varphi_{a,0}
-
\dfrac{\theta_{a,1}-\theta_{a,0}}
{\theta_{1,1}-\theta_{0,0}}
(\varphi_{1,1}-\varphi_{0,0})
\right)$ \\
\hline

RR-PTE$(a)$ 
& $\dfrac{\theta_{a,1}}{\theta_{a,0}}
\dfrac{\theta_{0,0}}{\theta_{1,1}}$
& $\dfrac{\theta_{0,0}}{\theta_{1,1}\theta_{a,0}}
\left(
\varphi_{a,1}
-
\dfrac{\theta_{a,1}}{\theta_{a,0}}\varphi_{a,0}
+
\dfrac{\theta_{a,1}}{\theta_{0,0}}\varphi_{0,0}
-
\dfrac{\theta_{a,1}}{\theta_{1,1}}\varphi_{1,1}
\right)$ \\
\hline

OR-PTE$(a)$ 
& $\dfrac{\theta_{a,1}(1-\theta_{a,0})}
{(1-\theta_{a,1})\theta_{a,0}}
\cdot
\dfrac{(1-\theta_{1,1})\theta_{0,0}}
{(1-\theta_{0,0})\theta_{1,1}}$
& $\text{OR-TE}\times \varphi(\text{OR-NIE}(a),P)$ \\
\hline

\end{tabular}

\end{table}





\paragraph{One-Step estimator} 
The One-Step estimator corrects the first-order bias $\mathbb{P}_n\{\varphi(O; \boldsymbol{ \widehat \eta}, \theta_{k,p}^{a,a^*}(\boldsymbol{ \widehat \eta}))\}$ by substracting it to the G-formula estimator. So, \(\widehat{\theta}_{k,p}^{a,a^*,\text{OS}}\) of \(\theta_{k,p}^{a,a^*}\) is defined as
$\widehat{\theta}_{k,p}^{a,a^*,\text{OS}}
=
\widehat{\theta}_{k,p}^{a,a^*}
+
\widehat{A}_{k,p}^{a,a^*}
+
\widehat{B}_{k,p}^{a,a^*},
$ where
\begin{align*}
\widehat{A}_{k,p}^{a,a^*}
&=
\frac{1}{n_0}
\sum_{\substack{i:\,A_i=a \\ S_i=k,\ \tilde{R}_i=0}}
\widehat{\Omega}_{k,p}^{a,a^*}(\boldsymbol{C}_i, M_i)
\Bigl\{
Y_i-\widehat{Q}^{a}_{k}(\boldsymbol{C}_i, M_i)
\Bigr\},
\\
\widehat{B}_{k,p}^{a,a^*}
&=
\frac{1}{n_0}
\sum_{\substack{i:\,A_i=a^* \\ S_i=p,\ R_i=0}}
\widehat{W}^{a^*}_{p}(\boldsymbol{C}_i)
\Bigl\{
\widehat{Q}^{a}_{k}(\boldsymbol{C}_i, M_i)
-
\widehat{b}_{k,p}^{a,a^*}(\boldsymbol{C}_i)
\Bigr\}.
\end{align*}

Here, \(\widehat{\theta}_{k,p}^{a,a^*}\) denotes the initial G-computation estimator of \(\theta_{k,p}^{a,a^*}\), as described in Section~2.4, but with the nuisance functions \(Q_{k}^{a}\) and \(b_{k,p}^{a,a^*}\) possibly estimated using flexible data-adaptive methods. Similarly, \(\widehat{\Omega}_{k,p}^{a,a^*}\) and \(\widehat{W}^{a^*}_{p}\) denote estimators of \(\Omega_{k,p}^{a,a^*}\) and \(W^{a^*}_{p}\), respectively, also obtained by using machine learning or other nonparametric estimation procedures.  

The OS estimator is \emph{doubly robust}, in the sense that
$\widehat{\theta}_{k,p}^{a,a^*\text{OS}}
=
\theta_{k,p}^{a,a^*}
+
o_p(1)
$, provided that either the outcome regression functions \({Q}_{k}^{a}\) and \({b_{k,p}^{a,a^*}}\), \emph{or} the nuisance parameters involved in the weight functions \({\Omega_{k,p}^{a,a^*}}\) and \({W}_{p}^{a^*}\), are consistently estimated. 
This property substantially reduces the risk of model misspecification relative to the fully parametric approaches described in the previous section. A proof of the double robustness property is provided in the Supplementary Materials.



\paragraph{TMLE estimator}

The TMLE procedure fluctuates the initial nuisance parameter estimates $\widehat{\boldsymbol{\eta}}$ along a least favorable submodel to obtain updated estimates $\widehat{\boldsymbol{\eta}}^{\mathrm{TMLE}}$ satisfying
$\mathbb{P}_n\{\varphi(O;\widehat{\boldsymbol{\eta}}^{\mathrm{TMLE}})\} = 0.$ In the present setting, this requires neutralizing both terms $\widehat{A}_{k,p}^{a,a^*}$ and $\widehat{B}_{k,p}^{a,a^*}$. We first update the initial estimator $\widehat Q_k^a$ through the fluctuation submodel
$\widehat Q_{k}^{a,(1)}
=
\widehat Q_k^a
+
\alpha \widehat{\Omega}_{k,p}^{a,a^*},$ where $\alpha$ is estimated by maximum likelihood. The corresponding score equation is exactly equal to the empirical bias term $\widehat{A}_{k,p}^{a,a^*}$, so that the resulting update removes this component of the bias. Using the updated regression function $Q_{k}^{a,(1)}$, we then construct an updated estimator $\widehat b_{k,p}^{a,a^*,(1)}$ of $\mathbb{E}\bigl[ Q_{k}^{a,(1)} \mid A=a^*, S=p, C\bigr].$
A second targeting step is subsequently performed via
$b_{k,p}^{a,a^*,(2)}
=\widehat b_{k,p}^{a,a^*,(1)}
+
\delta \widehat W_p^{a^*},
$
where $\delta$ is again estimated by maximum likelihood. By construction, the associated score equation corresponds to $\widehat{B}_{k,p}^{a,a^*}$, thereby eliminating the second empirical bias term. The Algorithm~\ref{tmle_algo} provides a detailed description of this procedure.

\begin{algorithm}[t]
\caption{Two-Step TMLE Targeting Algorithm}
\label{tmle_algo}
\begin{algorithmic}[1]
\Require $\{(Y_i, C_i, M_i, A_i, S_i, R_i, \tilde{R}_i)\}_{i=1}^n$; initial estimators
$\widehat{Q}_{k}^{a}$, $\widehat{b}_{k,p}^{a,a^*}$,
$\widehat{\Omega}_{k,p}^{a,a^*}$, $\widehat{W}^{a^*}_{p}$
\Ensure $\widehat{\theta}^{\mathrm{TMLE}}$

\State Obtain $\widehat{\alpha}$ by solving:
$$
\sum_{\substack{i:\,A_i=a \\ S_i=k,\ \tilde{R}_i=0}}
\widehat{\Omega}_{k,p}^{a,a^*}(\boldsymbol{C}_i, M_i)
\Bigl[
  Y_i - \widehat{Q}_{k}^{a}(\boldsymbol{C}_i, M_i)
  - \alpha\,\widehat{\Omega}_{k,p}^{a,a^*}(\boldsymbol{C}_i, M_i)
\Bigr] = 0.
$$
This is achieved, for instance, by fitting a linear regression model via maximum likelihood, among subjects with $A=a$, $S=k$, and $\tilde{R}=0$, using $Y$ as the outcome, $\widehat{Q}_{k}^{a}$ as an offset, and $\widehat{\Omega}_{k,p}^{a,a^*}$ as a covariate.
\State Obtain an updated estimate $\widehat{Q}_{k}^{a,(1)}$ of $Q_{k}^{a}$:
$$
\widehat{Q}_{k}^{a,(1)}(\boldsymbol{C}, \boldsymbol{M})
= \widehat{Q}_{k}^{a}(\boldsymbol{C}, \boldsymbol{M})
+ \widehat{\alpha}\,\widehat{\Omega}_{k,p}^{a,a^*}(\boldsymbol{C}, \boldsymbol{M}).
$$

\State Obtain an updated estimate $\widehat{b}^{a,a^{*},(1)}_{k,p}$ for $b^{a,a^{*}}_{k,p}$, by regressing $\widehat{Q}_{k}^{a,(1)}(\boldsymbol{C}, \boldsymbol{M})$ on $C$ among subjects with $A=a^*,S=p,$ and $R=0$.

\State Estimate $\widehat{\delta}$ by solving:
$$
\sum_{\substack{i:\,A_i=a \\ S_i=p,\ R_i=0}}
\widehat{W}^{a^*}_{p}(\boldsymbol{C}_i)
\Bigl[
  \widehat{Q}_{k}^{a,(1)}(\boldsymbol{C}_i, M_i)
  - \widehat{b}_{k,p}^{a,a^{*},(1)}(\boldsymbol{C}_i)
  - \delta\,\widehat{W}^{a^*}_{p}(\boldsymbol{C}_i)
\Bigr] = 0.
$$
This is achieved, for instance, by fitting a linear regression model via maximum likelihood, among subjects with $A=a$, $S=p$, and $R=0$, using $\widehat{Q}_{k}^{a,(1)}(\boldsymbol{C}, \boldsymbol{M})$ as the outcome, $\widehat{b}_{k,p}^{a,a^{*},(1)}$ as an offset, and $\widehat{W}^{a^*}_{p}$ as a covariate.
\State Obtain an updated estimate $\widehat{b}^{a,a^{*},(2)}_{k,p}$ of ${b}^{a,a^{*}}_{k,p}$:
$$
\widehat{b}^{a,a^{*},(2)}_{k,p}(\boldsymbol{C})
= \widehat{b}^{a,a^{*},(1)}_{k,p}(\boldsymbol{C})
+ \widehat{\delta}\,\widehat{W}^{a^*}_{p}(\boldsymbol{C}).
$$

\Return $\displaystyle
\widehat{\theta}^{\mathrm{TMLE}}
= \frac{1}{n_0}\sum_{i:\,S_i=0} \widehat{b}^{a,a^{*},(2)}_{k,p}(\boldsymbol{C}_i).
$
\end{algorithmic}
\end{algorithm}

\paragraph{Asymptotic properties of OS and TMLE estimators}  Let $\hat\theta_{k,p}^{a,a^*}$ denote the OS or TMLE estimator of $\theta_{k,p}^{a,a^*}$. Assume
(i) Positivity, described as identification Assumption ~\eqref{ass:positivity}; (ii) The second-order term $R(\boldsymbol{\hat\eta}, \bm \eta)$ is $o_P(n^{-1/2})$ and (iii) The class of functions $\{ o \rightarrow \varphi_{k,p}^{a,a^*}(o;\bm\eta',\theta'): |\theta'-\theta_0| < \delta, ||\bm \eta' - \bm \eta|| < \delta\}$ is Donsker for some $\delta>0$ and such that $P[\{\varphi_{k,p}^{a,a^*}(O;\bm \eta',\theta') - \varphi_{k,p}^{a,a^*}(O;\bm \eta,\theta_0)\}^2] \rightarrow 0$. In that case,
$\hat\theta_{k,p}^{a,a^*} = \theta_0
+ \frac{1}{n} \sum_{i=1}^n \varphi_{k,p}^{a,a^*}(O_i; \boldsymbol{\eta},   \theta_0) + o_P(n^{-1/2}),$
due to which $\sqrt{n}(\hat\theta_{k,p}^{a,a^*} - \theta_0) \xrightarrow{D} N(0,\sigma_{k,p}^{a,a^{*}\,2})$, where $\sigma_{k,p}^{a,a^{*}\,2} = V\{ \varphi_{k,p}^{a,a^*}(O; \bm \eta, \theta_0)\}$ is the non parametric efficiency bound.

Note that condition (ii) for asymptotic normality is satisfied if all components of $\hat{\eta}$ converges in $L_2(P)$ norm to their true counterparts in $\eta$ at $n^{-1/4}$-rate or faster. This is the case for many data-adaptive algorithms such as LASSO or highly adaptive LASSO, under certain conditions \citep{benkeser2016highly}. In contrast, condition (iii) (i.e. Donsker condition) may be avoided by using cross-fitting in the estimation procedure. To achieve this, the dataset is randomly partitioned into $Q$ sets of approximately equal size, namely $D_1, \ldots , D_Q$. On each sample $T_q = \{1,\ldots,n\} \setminus D_q$, the data-adaptive algorithm will be trained and then used to produce a prediction $\hat{\eta}^*$ of $\eta$ for each patient in the validation set $V_q$. The One-Step and TMLE estimators are finally adapted to cross-fitting by substituting all occurrences of $\boldsymbol{\hat{\eta}}( O_i)$ by $\boldsymbol{\hat{\eta}}^{*}( O_i)$ in the estimation procedure. 

As a direct consequence of the above asymptotic result, the variance of the estimators $\widehat \theta_{k,p}^{a,a^*, \text{OS}}$, $\widehat \theta_{k,p}^{a,a^*, \text{TMLE}}$ can be estimated by the sample variance of the EIF, i.e. $\widehat  \sigma_{k,p}^{a,a^*, 2} = \hat V(\varphi_{k,p}^{a, a^*}(O;\boldsymbol{\hat \eta},\hat \theta_{k,p}^{a,a^*}))$, with $\widehat\theta_{k,p}^{a,a^*}$ and the nuisance parameter vector $\boldsymbol{\hat \eta}$ estimated as described above. In what follows, an estimator for $\widehat \zeta_{k,p} = \widehat \theta_{k,p}^{a,a} - \widehat \theta_{k,p}^{a,a^*}  $ and its variance can be constructed from $\widehat \theta_{k,p}^{a,a},\widehat \theta_{k,p}^{a,a^*} $ and $\widehat  \sigma_{k,p}^{a,a},\widehat \sigma_{k,p}^{a,a^*} $ by applying the Delta method, as is done for the simple G-computation approach. 

\subsection{Heterogeneity tests}
To assess mediator-related heterogeneity, one may perform a Wald test of the composite null hypothesis that, for every outcome source $k$, the standardized effects are identical across mediator sources, i.e.:
$$H_0: \zeta_{k,1} = \ldots = \zeta_{k,P} \quad \forall k\in \{1,\ldots,K\}.$$
For each outcome source $k$, this hypothesis can be expressed as a set of $P-1$ linear restrictions,$$
\begin{pmatrix}
\zeta_{k,1} - \zeta_{k,2} &
\zeta_{k,1} - \zeta_{k,3} &
\ldots &
\zeta_{k,1} - \zeta_{k,P}
\end{pmatrix}^\top
=
\mathbf{0}_{P-1}.
$$

Collecting all restrictions across outcome sources yields the composite hypothesis $H_0: \boldsymbol{G}\boldsymbol{\zeta}=\mathbf{0},
$ where $\boldsymbol{G}$ is a contrast matrix of dimension $K(P-1)\times KP$. The Wald statistic is $$
W=
(\boldsymbol{G}\hat{\boldsymbol{\zeta}})^\top
\left(
\boldsymbol{G}\widehat{\Sigma}\boldsymbol{G}^\top
\right)^{-1}
(\boldsymbol{G}\hat{\boldsymbol{\zeta}}).
$$
where $\hat{\Sigma}$ is the (asymptotic) covariance matrix of $\boldsymbol{\hat \zeta}$. 
Under $H_0$ and standard regularity conditions, $W{\sim}
\chi^2_{K(P-1)}.$ A significant test indicates that at least one standardized effect $\zeta_{k,p}$ differs across all $\zeta_{k,p}$ sharing the same outcome component $p$, which provides evidence of mediator-related heterogeneity.

Similarly, outcome-related heterogeneity may be assessed by testing whether, for every mediator source $p$, the standardized effects are identical across outcome sources; that is,
$$H_0^*: \zeta_{1,p} = \ldots = \zeta_{K,p} \quad \forall p\in \{1,\ldots,P\}.$$ 
In practice, however, measures that quantify the magnitude of heterogeneity are generally preferred over formal hypothesis tests, as the assumption of no heterogeneity is often unrealistic and therefore of limited scientific interest. Moreover, heterogeneity measures provide a more informative characterization of the extent to which study-specific effects vary across sources. We therefore develop such measures in the following section.

\section{Meta-analysis of standardized effect estimates}

\subsection{A novel random-effect meta-analysis model}To summarized the standardized effect estimates $\widehat{\boldsymbol{\zeta}}=\{\widehat{\zeta}_{k,p}: k=1,\ldots,K, p=1,\ldots,P\}$, we posit the following random-effects model:
\begin{equation}\label{eq:model}
    \widehat{\zeta}_{k,p}
    = \zeta + \gamma_k + \alpha_p + \varepsilon_{k,p},
\end{equation}
where $\zeta$ denotes the summary effect, $\varepsilon_{k,p}$ represents sampling variability arising from estimation of $\zeta_{k,p}$, while $\gamma_k$ and $\alpha_p$ are random effects that account for clustering induced by the mediator source and outcome source, respectively. The random components ${\varepsilon_{k,p},\gamma_k,\alpha_p}$ are assumed to be pairwise independent. In addition, we assume that $\boldsymbol{\varepsilon}=(\varepsilon_{k,p})_{\substack{k=1,\ldots,K\\ p=1,\ldots,P}} \sim\mathcal{N}(0,\Sigma)$, where $\Sigma$ is a known unstructured covariance matrix. This assumption is justified by the asymptotic normality of $\widehat{\bm\zeta}$. Finally, $\gamma_k$ and $\alpha_p$ are assumed to be normal with mean zero and variances $\xi^2$ and $\eta^2$, respectively. 

Within each subset $\mathcal{S}_k=\{\zeta_{k,1},\zeta_{k,2},\ldots\}$, the variability across $\zeta_{k,p}$ is attributed to differences in the mediator distribution across source $p$. Under model~\eqref{eq:model}, this variability is invariant in $k$ and equals $\eta^2$, the so-called mediator-related heterogeneity variance. Likewise, the variability across elements in $\mathcal{S}_{\cdot p}=\{\zeta_{1,p},\zeta_{2,p},\ldots\}$ is invariant in $p$ and equals $\xi^2$, the so-called outcome-related heterogeneity variance.
The total covariance matrix $V$ of $\widehat{\bm \zeta}$ can then be decomposed into three components, including two heterogeneity variances and chance. Specifically,
\begin{equation}\label{eq:V}
    V
    = \xi^2\!\left(I_K \otimes \mathbf{1}_P\mathbf{1}_P^{\top}\right)
    + \eta^2\!\left(\mathbf{1}_K\mathbf{1}_K^{\top} \otimes I_P\right)
    + \Sigma,
\end{equation}
i.e.\ entry-wise, $V_{(k,p),(k',p')}
= \xi^2\,\mathbb{I}_{k=k'}
+ \eta^2\,\mathbb{I}_{p=p'}
+ \Sigma_{(k,p),(k',p')},$
where \( I_P \) is the identity matrix in \( \mathbb{R}^{P \times P} \), \( \mathbf{1}_P \) is the column vector of ones of size \( P \), and \( \otimes \) denotes the Kronecker product of matrices.


To estimate model~\eqref{eq:model}, one may consider restricted maximum likelihood (REML) estimation or Bayesian Markov chain Monte Carlo (MCMC) methods. REML, however, may encounter convergence difficulties due to the complex covariance structure. MCMC methods, by contrast, offer greater flexibility but may be sensitive to the choice of prior distributions. Furthermore, estimates of the summary effect and heterogeneity variances may be biased if the proposed random-effects model is misspecified, for example, if the two random effects, $\gamma_k$ and $\alpha_p$, are correlated. More fundamentally, random-effects models implicitly assume that the eligible studies constitute a random sample drawn from a hypothetical superpopulation of studies. This assumption is largely untestable and is often adopted primarily for statistical convenience.

\subsection{A non-parametric meta-analysis approach based on ANOVA decomposition}
To avoid the superpopulation assumption associated with (parametric) random-effect models, one can directly summarize $\zeta_{k,p}$ by specifying weight vectors $(w_i)_{i=1}^K$ and $(q_i)_{i=1}^P$ for outcome and mediator sources, respectively, such that $\sum_i w_i = 1$ and $\sum_i q_i = 1$. A simple choice is uniform weighting, i.e., $w_k = 1/K$ and $q_p=1/P$. Alternatively, one may choose $w_k=n_k/\sum_{k=1}^K n_k$ and $q_p=m_p/\sum_{p=1}^P m_p$, where $n_k$ and $m_p$ denote the sample sizes corresponding to outcome source $k$ and mediator source $p$, respectively. This weighting scheme makes the contribution of each data source proportional to its sample size.  Importantly, using simple weighting schemes rather than inverse-variance weights,
as in standard meta-analysis, can prevent
complications in establishing the asymptotic behaviors of the summary effect and heterogeneity variance estimates (see below). 

Under a prespecified weighting scheme, the summary effect across studies is defined as the weighted average of all standardized effects $\zeta_{k,p}$, i.e. $\zeta = \sum_{k,p} w_kq_p\zeta_{k,p}$. 
In the absence of heterogeneity, this summary effect reduces to the natural indirect effect in the target population. To estimate $\zeta$, one can simply plug in the estimates $\hat\zeta_{k,p}$ of $\zeta_{k,p}$ proposed in the previous section. The asymptotic variance of $\hat\zeta$ can then be established by using Delta method. 

In what follows, we define the total variability across studies as the following (weighted) variance:
$$\tau^2 = \sum_{k,p}w_kq_p(\zeta_{k,p} - \zeta)^2,$$
which takes null value in the absence of heterogeneity. The classical ANOVA sum-of-squares decomposition then implies that:
\begin{equation} \label{eq:anova}
    \tau^2 = \underbrace{\sum_k w_k(\zeta_{k,\cdot} - \zeta)^2}_{\text{Outcome-related variation}} + \underbrace{\sum_p q_p(\zeta_{\cdot,p} - \zeta)^2}_{\text{Mediator-related variation}} + \underbrace{\sum_{k,p}w_kq_p(\zeta_{k,p}-\zeta_{k,\cdot}-\zeta_{\cdot,p}+\zeta)^2}_{\text{Interaction}}.
\end{equation}
\\
The first component on the right-hand side measures the variability of the $k$-specific averages $\zeta_{k,\cdot}$ around the overall mean $\zeta$. This component equals zero whenever $\zeta_{k,p}=\zeta_{k',p}$ for all $k,k',p$, indicating the absence of outcome-related heterogeneity. Similarly, the second term represents the variability of the $p$-specific averages $\zeta_{\cdot,p}$ around the overall mean $\zeta$. This component equals zero whenever $\zeta_{k,p}=\zeta_{k,p'}$ for all $k,p,p'$, indicating the absence of mediator-related heterogeneity. The final component captures variability not attributable to either source alone. Such a component will equal zero whenever the variability across elements in $\boldsymbol{S_k} = \{\zeta_{k,1},\ldots, \zeta_{k,P}\}$ and $\boldsymbol{S_p} = \{\zeta_{1,p},\ldots, \zeta_{K,p}\}$ are invariant in $k$ and $p$, respectively. All components can be estimated by plugging in the estimates $\hat\zeta$ of $\zeta$ obtained previously.

\begin{remark} \label{remark:anova}
To connect the proposed ANOVA decomposition with the random-effects model in \eqref{eq:model}, note that any collection of standardized effects $\{\zeta_{k,p}\}$ can be uniquely decomposed as:
\[
\zeta_{k,p}
=
\zeta
+
\gamma_k
+
\alpha_p
+
\delta_{k,p},
\]
where
$\gamma_k=\zeta_{k,\cdot}-\zeta;
\alpha_p=\zeta_{\cdot,p}-\zeta
$ and $\delta_{k,p}
=
\zeta_{k,p}
-
\zeta_{k,\cdot}
-
\zeta_{\cdot,p}
+
\zeta$
are the ANOVA effects satisfying the weighted constraints:
\[
\sum_{k=1}^K w_k\gamma_k=
\sum_{p=1}^P q_p\alpha_p=\sum_{k=1}^K w_k\delta_{k,p}=\sum_{p=1}^P q_p\delta_{k,p}=0.
\]
This decomposition is purely algebraic and does not require any distributional assumptions. Under the additional \emph{additivity assumption} that 
$\delta_{k,p}=0 ~ \forall k,p$,
the decomposition reduces to:
\[
\zeta_{k,p}
=
\zeta
+
\gamma_k
+
\alpha_p,
\]
which can be viewed as the finite-population analogue of the random-effects model \eqref{eq:model}. Thus, the interpretation of the total heterogeneity as the sum of outcome- and mediator-related components in this model implicitly relies on the absence of interaction between outcome and mediator sources.
\end{remark}

\section{Simulations}

We conducted a simulation study to evaluate the finite-sample performance of the proposed estimators. In each simulation replicate, we generated data from one target population and five trial populations, yielding a total sample size of 20,000 observations. In Scenario 1, the data generation mechanism is presented as follows:
\noindent
\begin{align*}
C_1 &\sim \mathcal{U}(0,1), \quad C_2 \sim \mathcal{B}(0.5) \\
\mathbb{P}(S=s\mid C_1,C_2)
&= \mathbb{P}(S=0\mid C_1,C_2)\exp(  (C_1, C_2) \bm \gamma_s ), \quad s = 1, \ldots,5 \\
\mathbb{P}(S=0\mid C_1,C_2)
&=\left[1+\sum_{s=1}^5 \exp((C_1, C_2) \bm \gamma_s)\right]^{-1} \\
\mathbb{P}(A=1\mid C_1, C_2,S) &=0.5\\
\mathbb{P}(Y=1\mid A,M,C_1, C_2,S)
&=\mathrm{expit}\Big(
-8\,\mathbb{I}(0.2<C_1\le0.6)
+0.5C_2+2M+MC_2+AC_2+\beta_s A
\Big) \\
\mathbb{P}(M=1\mid A,C_1, C_2,S)
&=\mathrm{expit}\Big(
-1-2C_1\,\mathbb{I}(0.2<C_1\le0.6)
+C_2+\omega_s A
\Big), 
\end{align*}
with the following coefficients:
\begin{align*}
\bm\gamma_{1}=\bm\gamma_{2}= \bm\gamma_{3}=(0.15,-0.10,-0.10)^\top,&\quad
\bm\gamma_{4}=\bm\gamma_{5}=(-0.15,0.10,0.10)^\top \\
\bm\omega_{0}=\bm\omega_{5}=(0.5,0.8,0.5,1.2)^\top,\quad
\bm\omega_{1}=&\bm\omega_{3}=(0.5,0.8,0.5,1.4)^\top,\quad
\bm\omega_{2}=\bm\omega_{4}=(0.5,0.8,0.5,1.8)^\top.
\end{align*}

In Scenario 2, we adopted the same data generating mechanism but introduced missing data in the mediator and outcome by the following models:
\begin{align*}
    \mathbb{P}(R=0\mid A,C_1, C_2,S)
&=\mathrm{expit}(0.1A-C_1-C_2S-\sqrt{C_2}) \\
\mathbb{P}(\tilde{R}=0\mid A,C_1, C_2,M,S)
&=\mathrm{expit}(0.5\,\mathbb{I}_{S=1}+0.1A-C_1-C_2S+0.05M).
\end{align*}

Across both scenarios, we compared parametric estimators (G-formula and the two version of inverse probability weighting called IPW (Equation~\eqref{IPW_1}) and IPW 2 (Equation~\eqref{IPW_2})) with semiparametric estimators OS and TMLE. For the parametric estimators, all nuisance functions were estimated using logistic regression except $b_{k,p}^{a,a^*}$, which was estimated by using linear regression.

For OS and TMLE, we considered two nuisance-estimation strategies. The first strategy (OS GLM and TMLE GLM) used parametric methods as above. In the second strategy (OS RF and TMLE RF), random forests from the \texttt{Python} library \texttt{scikit-learn} were used to estimate all nuisance functions, except $b_{k,p}^{a,a^*}$, which was estimated by using honest forests regression from \texttt{econml}.

In the meta-analysis step, the standardized effect estimates are summarized by the ANOVA-based method proposed in Section 3.2, using a uniform weighting scheme. 

For each scenario, we report relative bias, empirical variance, estimated variance, and coverage of the 95\% confidence interval of $\hat\zeta_{k,p}$ and the summary effect estimate $\hat\zeta$. For $\hat\zeta_{k,p}$ (and likewise for $\hat\zeta)$, relative bias (in \%) is defined as
$100 \times (\widehat{\zeta}_{k,p}-\zeta_{k,p})/\zeta_{k,p},$ and averaged over simulation replicates. Empirical variance, $\operatorname{Var}(\widehat{\zeta}_{k,p})$, is the Monte Carlo variance of the estimator across replicates, while $\widehat{\operatorname{Var}}(\widehat{\zeta}_{k,p})$ is the median of the estimated variances across simulations. Coverage is the proportion of replicates in which the nominal 95\% confidence interval contains $\zeta_{k,p}$. To assess the performance of the proposed heterogeneity decomposition, we report the median estimated value of each heterogeneity component across simulation replicates and its corresponding true value.

\begin{table}[ht]
\centering
\caption{Performance of estimators under outcome and mediator misspecification, under scenarios with and without missing data. RF denotes Random Forest-based nuisance estimation, and GLM denotes generalized linear models. Performances are averaged over the 25 estimators $\{\widehat{\zeta}_{k,p}\}_{k,p}$. $\operatorname{Var}(\widehat{\zeta}_{k,p})$ denotes the empirical variance over the Monte Carlo replications and $\widehat{\operatorname{Var}}(\widehat{\zeta}_{k,p})$ denotes the median of the estimated variances across simulations.}
\label{table:simus}

\resizebox{\textwidth}{!}{%
\begin{tabular}{lcccccccc}
\toprule
& \multicolumn{4}{c}{No missing data}
& \multicolumn{4}{c}{Missing data} \\
\cmidrule(lr){2-5}
\cmidrule(lr){6-9}
& Bias (\%) & $\widehat{\operatorname{Var}}(\widehat{\zeta}_{k,p})$ & $\operatorname{Var}(\widehat{\zeta}_{k,p})$ & Coverage
& Bias (\%) & $\widehat{\operatorname{Var}}(\widehat{\zeta}_{k,p})$ & $\operatorname{Var}(\widehat{\zeta}_{k,p})$ & Coverage \\
\midrule

OS RF 
& $4.81\times10^{-1}$ & $3.00\times10^{-5}$ & $3.40\times10^{-5}$ & $93.3\%$
& $6.96\times10^{0}$ & $5.70\times10^{-5}$ & $6.70\times10^{-5}$ & $90.5\%$ \\

OS GLM 
& $1.16\times10^{2}$ & $6.40\times10^{-5}$ & $6.40\times10^{-5}$ & $0.2\%$
& $1.12\times10^{2}$ & $1.03\times10^{-4}$ & $1.01\times10^{-4}$ & $2.1\%$ \\

TMLE RF 
& $3.90\times10^{0}$ & $3.10\times10^{-5}$ & $2.60\times10^{-5}$ & $94.7\%$
& $1.14\times10^{1}$ & $1.79\times10^{-4}$ & $2.09\times10^{-4}$ & $91.4\%$ \\

TMLE GLM
& $1.17\times10^{2}$ & $6.40\times10^{-5}$ & $6.70\times10^{-5}$ & $0.2\%$
& $1.26\times10^{2}$ & $1.02\times10^{-4}$ & $9.00\times10^{-5}$ & $0.6\%$ \\

IPW GLM 
& $1.20\times10^{2}$ & $3.02\times10^{-4}$ & $3.65\times10^{-4}$ & $32.5\%$
& $1.16\times10^{2}$ & $4.44\times10^{-4}$ & $4.47\times10^{-4}$ & $46.9\%$ \\

IPW2 GLM 
& $1.15\times10^{2}$ & $3.76\times10^{-4}$ & $3.69\times10^{-4}$ & $41.4\%$
& $1.22\times10^{2}$ & $8.19\times10^{-4}$ & $4.03\times10^{-4}$ & $67.9\%$ \\

G-formula GLM
& $1.16\times10^{2}$ & $6.70\times10^{-5}$ & $6.70\times10^{-5}$ & $0.2\%$
& $1.27\times10^{2}$ & $9.60\times10^{-5}$ & $9.00\times10^{-5}$ & $0.5\%$ \\

\bottomrule
\end{tabular}%
}
\end{table}

Results of this simulation study are presented in Tables 2 and 3. In Table 2, each performance metric is first computed for each of the 25 estimands $\zeta_{k,p}$ and then averaged across all estimands. Overall, the parametric estimators perform poorly due to model misspecification. In contrast, the OS RF and TMLE RF approaches exhibit satisfactory performance when the nuisance functions are estimated using data-adaptive methods. Although the performance of both approaches deteriorates in the presence of missing mediator and outcome data, the empirical coverage of the corresponding 95\% CI remains above 90\%, indicating adequate finite-sample performance despite the additional missingness.

Table \eqref{tab:anova_decomp_missing_comparison} reports the median estimated heterogeneity components across simulation replicates, together with their corresponding oracle values. As shown in the table, outcome-related heterogeneity is the primary contributor to the variability across the standardized effects $\zeta_{k,p}$, whereas mediator-related heterogeneity and the interaction component are negligible. This pattern is accurately recovered when the standardized effects are estimated consistently, for example using the OS RF or TMLE RF estimators.

One limitation is that when the true value of a heterogeneity component is close to zero, that is, on the boundary of the parameter space, its estimator may exhibit finite-sample bias. This phenomenon is observed for the mediator-related heterogeneity $\eta^2$ in our simulation setting. Overall, TMLE RF yields smaller bias when estimating $\eta^2$ than OS RF, although the presence of missing mediator and outcome data increases the bias for both estimators. Nevertheless, these biases are sufficiently small that they do not affect the qualitative conclusion regarding the relative importance of the different sources of heterogeneity.

Finally, Table \eqref{table:simus_summary} reports summary statistics on estimator performance, including bias, variance, and coverage, for the aggregated summary $\widehat{\zeta}^{a=1}$. RF-based estimators (One-Step RF and TMLE RF) achieve substantially lower bias and near-nominal coverage, whereas linear-based estimators exhibit a much larger bias, driving coverage down to zero.

\begin{table}[ht]
\centering
\caption{Median ANOVA decomposition terms ($\eta^2$, $\xi^2$, and interaction) across 500 simulations, by estimator and nuisance model, under scenarios with and without missing data. The oracle values are computed from the true $\zeta_{k,p}$ values. RF denotes Random Forest-based nuisance estimation, and GLM denotes generalized linear models.}
\label{tab:anova_decomp_missing_comparison}
\scalebox{0.85}{%
\begin{tabular}{l|ccc|ccc}
\toprule
& \multicolumn{3}{c|}{No missing data} & \multicolumn{3}{c}{Missing data} \\
& $\eta^2$ & $\xi^2$ & Interaction
& $\eta^2$ & $\xi^2$ & Interaction \\
\midrule
Oracle
& $1.08\times10^{-5}$ & $8.51\times10^{-5}$ & $7.49\times10^{-18}$
& $1.08\times10^{-5}$ & $8.51\times10^{-5}$ & $7.49\times10^{-18}$ \\

\midrule
OS RF
& $2.09\times10^{-5}$ & $9.14\times10^{-5}$ & $0.00$
& $3.92\times10^{-5}$ & $8.69\times10^{-5}$ & $1.39\times10^{-19}$ \\

OS GLM
& $2.58\times10^{-5}$ & $4.16\times10^{-4}$ & $0.00$
& $1.59\times10^{-4}$ & $4.07\times10^{-4}$ & $0.00$ \\

TMLE RF
& $1.91\times10^{-5}$ & $8.26\times10^{-5}$ & $-2.78\times10^{-19}$
& $3.95\times10^{-5}$ & $1.53\times10^{-4}$ & $0.00$ \\

TMLE GLM
& $2.66\times10^{-5}$ & $4.30\times10^{-4}$ & $0.00$
& $1.70\times10^{-4}$ & $4.66\times10^{-4}$ & $0.00$ \\

IPW GLM
& $4.74\times10^{-5}$ & $6.05\times10^{-4}$ & $0.00$
& $2.07\times10^{-4}$ & $6.25\times10^{-4}$ & $0.00$ \\

IPW2 GLM
& $2.64\times10^{-5}$ & $6.68\times10^{-4}$ & $0.00$
& $1.06\times10^{-4}$ & $6.58\times10^{-4}$ & $-1.11\times10^{-18}$ \\

G-formula GLM
& $2.70\times10^{-5}$ & $4.29\times10^{-4}$ & $1.11\times10^{-18}$
& $1.59\times10^{-4}$ & $4.81\times10^{-4}$ & $1.11\times10^{-18}$ \\

\bottomrule
\end{tabular}%
}
\end{table}

\begin{table}[ht]
\centering
\caption{Performance of the summary estimator $\widehat{\zeta}$ across estimators under scenarios with and without missing data.}
\label{table:simus_summary}
\resizebox{\textwidth}{!}{%
\begin{tabular}{lccc ccc}
\toprule
& \multicolumn{3}{c}{No missing data}
& \multicolumn{3}{c}{Missing data} \\
\cmidrule(lr){2-4}
\cmidrule(lr){5-7}
& Bias (\%) & $\widehat{\operatorname{Var}}(\widehat{\zeta})$ & Coverage
& Bias (\%) & $\widehat{\operatorname{Var}}(\widehat{\zeta})$ & Coverage \\
\midrule
OS RF
& $3.41\times10^{-1}$ & $5.00\times10^{-6}$ & $92.8 $
& $4.21\times10^{0}$ & $8.00\times10^{-6}$ & $86.6$ \\

OS GLM
& $1.16\times10^{2}$ & $1.20\times10^{-5}$ & $0.0$
& $1.12\times10^{2}$ & $1.90\times10^{-5}$ & $0.0$ \\

TMLE RF
& $3.91\times10^{0}$ & $5.00\times10^{-6}$ & $90.0$
& $1.12\times10^{1}$ & $3.20\times10^{-5}$ & $85.6 $ \\

TMLE GLM
& $1.16\times10^{2}$ & $1.20\times10^{-5}$ & $0.0 $
& $1.25\times10^{2}$ & $1.90\times10^{-5}$ & $0.0 $ \\

IPW GLM
& $1.19\times10^{2}$ & $8.80\times10^{-5}$ & $0.0 $
& $1.16\times10^{2}$ & $8.50\times10^{-5}$ & $0.0$ \\

IPW2 GLM
& $1.14\times10^{2}$ & $7.60\times10^{-5}$ & $0.0 $
& $1.21\times10^{2}$ & $1.32\times10^{-4}$ & $0.0 $\\

G-formula GLM
& $1.15\times10^{2}$ & $1.30\times10^{-5}$ & $0.0 $
& $1.26\times10^{2}$ & $1.80\times10^{-5}$ & $0.0 $ \\

\bottomrule
\end{tabular}%
}
\end{table}

\section{Application on Real Data}

To illustrate the proposed methodology, we investigate the mediating role of self-reported health in the relationship between educational attainment and subjective well-being, using data across different countries from the 2017--2021 World Values Survey, Wave 7, Master Survey Questionnaire \citep{haerpfer2020world}. The core model considers three variables: the exposure $A$, defined as attainment of higher education (coded as a binary indicator); the mediator $M$, representing self-reported health, originally measured on a five-point ordinal scale (0--4); and the outcome $Y$, denoting overall life satisfaction, measured on a ten-point ordinal scale (1--10). To facilitate estimation, the mediator was dichotomized. Categories 0, 1, and 2 (indicating poor to moderate self-reported health) were regrouped into $M=0$ (``poor health''), while categories 3 and 4 were regrouped into $M=1$ (``good health''). 

\paragraph{Country Selection}

To enable a meaningful comparative analysis while avoiding excessive fragmentation of the data, we restricted the initial pool of 60 countries surveyed in Wave 7 to a smaller, purposively selected subset of 10 countries: Australia, Canada, Germany, Mongolia, Netherlands, Russia, Slovakia, Ukraine, United Kingdom and United States of America. The chosen target country is Canada. This selection was guided by four criteria: (i) cross-national diversity, prioritizing countries with markedly different health and education systems in order to capture heterogeneity in the exposure--mediator--outcome relationships across institutional contexts; (ii) sample size, favoring countries with larger sample to maximize statistical power and yield more precise parameter estimates by reducing standard errors; (iii) data completeness, excluding countries for which one or more required covariates were entirely unavailable, so as to avoid the systematic exclusion of all observations from a given country during complete-case analysis; and (iv) \textit{positivity}, retaining only countries for which every category defined by the covariates exhibited a non-zero probability of exposure and mediator, thereby ensuring the identifiability of the causal mediation parameters of interest. The corresponding positivity diagnostics for each retained country are reported in Appendix~\eqref{ass:positivity}.

\paragraph{Covariates}
We adjusted for six sociodemographic confounders commonly available across WVS countries: age, employment status, sex, religion, degree of urbanization, and marital status. Age is included as a continuous variable (in years). Employment status is coded as a binary indicator distinguishing individuals currently in paid employment from those who are not (e.g., unemployed, retired, students, or homemakers). Sex is coded as a binary variable (male/female). Self-reported religiosity 
is coded on a 1–4 scale. Degree of urbanization is coded as a binary variable distinguishing urban from rural areas of residence. Finally, marital status is coded as a binary indicator distinguishing married (or cohabiting) respondents from all other categories (single, divorced, separated, or widowed). 

\paragraph{Results} Table~\ref{tab:data_results} reports the estimation results obtained using the OS and TMLE approaches. For the nuisance parameter estimation, we employ honest random forests from \texttt{econml} for the outcome regression models and L2-penalized logistic regression, implemented using \texttt{scikit-learn}, for the propensity score models.

\begin{table}[ht]
\centering
\resizebox{\textwidth}{!}{%
\begin{tabular}{l|cc|ccc|ccc}
\toprule
 & \multicolumn{2}{c|}{Aggregate} & \multicolumn{3}{c|}{Heterogeneity} & \multicolumn{3}{c}{Contribution of $\tau^2$ (\%)} \\
\cmidrule(lr){2-3} \cmidrule(lr){4-6} \cmidrule(lr){7-9}
 & $\hat{\zeta}$ (95\% CI) & $\hat{V}$ & $\hat{\eta}^2$ & $\hat{\xi}^2$ & Interaction  & Mediation & Outcome & Interaction \\
\midrule
OS
& $8.243 \times 10^{-2} [8.192 \times 10^{-2}, 8.294 \times 10^{-2}]$
& $2.599 \times 10^{-4}$
& $1.939 \times 10^{-3}$
& $3.557 \times 10^{-3}$
& $3.608 \times 10^{-18}$
& $33.69$
& $61.80$
& $4.11 \times 10^{-13}$ \\

TMLE
& $9.310 \times 10^{-2} [9.259 \times 10^{-2}, 9.361 \times 10^{-2}]$
& $2.599 \times 10^{-4}$
& $2.669 \times 10^{-3}$
& $3.355 \times 10^{-3}$
& $1.263 \times 10^{-17}$
& $42.47$
& $53.39$
& $1.27 \times 10^{-13}$ \\

\bottomrule
\end{tabular}%
}
\caption{Comprehensive estimation results: aggregate estimates and heterogeneity decomposition for OS and TMLE estimators. 95\% confidence intervals are shown in brackets for $\hat{\zeta}$ values.}
\label{tab:data_results}
\end{table}

Table ~\eqref{tab:data_results} reports the estimated aggregated natural indirect effect mediated through health in the relationship between educational attainment and life satisfaction on the additive scale. The analysis highlights substantial cross-country heterogeneity even after accounting for case-mix heterogeneity. This suggests that the mediated effect of higher education on life satisfaction is not uniform across societies, as the outcome and mediator mechanisms linking education to well-being vary across countries.

TMLE attributes $42.47\%$ of total variability to the  mediator-related heterogeneity ($\hat{\eta}^2_{\text{TMLE}} = 2.669 \times 10^{-3}$) versus $33.69\%$ for OS ($\hat{\eta}^2_{\text{OS}} = 1.939 \times 10^{-3}$). This finding suggests that cross-country differences in the mediator contribute less to the heterogeneity of the indirect effect than cross-country differences in life satisfaction.

\section{Conclusion}

In this work, we develop a novel two-stage approach in a defined target population framework, where standardized indirect effects are estimated by combining information on exposure–mediator and mediator–outcome relationships obtained from external eligible studies. We propose three parametric and two nonparametric estimators of the natural indirect effect on the risk difference scale, and derive their efficient influence functions under several causal estimands and scales. The asymptotic variance of these estimators is obtained from the empirical variance of the efficient influence function for the nonparametric estimators, and via M-estimation for the parametric estimators. 
We further introduce a non-parametric meta-analysis approach, based on ANOVA decomposition, that accounts for two distinct sources of variability: treatment–outcome heterogeneity arising from different versions of the treatment, and treatment–mediator heterogeneity. The performance of the proposed methods is assessed through simulation studies under challenging misspecification scenarios in finite samples. Finally, we apply our approach to real-world data to evaluate the heterogeneities in the natural indirect effect of higher education on life satisfaction through health across countries.

A key limitation of our methods is their reliance on access to individual participant data (IPD), which are often unavailable in real-world applications \citep{ahmed2012assessment}.The unavailability of IPD may introduce bias due to selective data availability. Future work will focus on extending this framework to accommodate studies without IPD, thereby enabling the incorporation of aggregate-level information and enhancing the utility of the proposed methods for evidence synthesis in a wider range of settings.

\section{Competing interests}
No competing interest is declared.

\section{Acknowledgments}
T.T.V is supported by the French National Research Agency (Agence Nationale de la Recherche),
through a funding for Chaires de Professeur Junior (23R09551S-MEDIATION).
\bibliographystyle{plainnat}
\bibliography{mybibilo.bib}

\appendix

\section{Identification} \label{proof_identification}
\begin{eqnarray*}
    \mathbb{E} \left[Y(a,k, M(a^{*}, p)) \mid j\right] &=&\mathbb{E}_{C} \left[\mathbb{E} \left[Y(a,k, M(a^{*}, p) \mid C \right] \mid j\right] \\
     && \\
    \textit{ Cross-World Ind.} \hookrightarrow   && \\ 
    &=&\mathbb{E}_{C} \left[\mathbb{E}_{M} \left[\mathbb{E}\left[Y(a,k, M(a^{*}, p) \mid C, M(a^{*}, p)\right ] \right] \mid j\right] \\    
    &=& \mathbb{E}_{C} \left[  \sum_{m  \in \mathcal{M}} \mathbb{E} \left[Y(a,k, M(a^{*}, p) \mid C, M(a^{*}, p) = m \right ] \right. \\
    &&\quad \mathbb{P}(M(a^{*}, p) = m \mid C) \mid j \biggr] \\    
     \textit{ Outcome Transportability} \hookrightarrow    \\ 
    &=&  \mathbb{E}_{C} \left[  \sum_{m  \in \mathcal{M}} \mathbb{E} \left[Y(a, k, M(a^{*}, p) \mid C, M(a^{*}, p) = m, k \right ] \right. \\
    &&  \quad \quad \mathbb{P}(M(a^{*}, p) = m \mid C) \mid j  \biggr] \\
    \textit{ Outcome Conditionnal } \hookrightarrow   \\
    &=&  \mathbb{E}_{C} \left[  \sum_{m  \in \mathcal{M}} \mathbb{E} \left[Y(a,k, M(a^{*}, p) \mid C, M(a^{*}, p) = m,  k, a \right ]  \right. \\
    && \left. \quad \mathbb{P}(M(a^{*}, p) = m \mid C) \mid j \right] \\  
     \textit{ Outcome Constistency  } \hookrightarrow 
    && \\
    &=&  \mathbb{E}_{C} \left[  \sum_{m  \in \mathcal{M}} \mathbb{E} \left[Y \mid C, M = m, a, k \right ] \right. \\
     &&  \quad \quad \mathbb{P}(M(a^{*}, p) = m \mid C) \mid j\biggr]  \\
      \textit{Mediator Transportability} \hookrightarrow  
          && \\
        &=&  \mathbb{E}_{C} \left[  \sum_{m  \in \mathcal{M}} \mathbb{E}  \left[Y \mid C, M = m, a, k \right ] \right. \\
        && \quad \mathbb{P}(M(a^{*}, p) = m \mid C,  p) \mid j \biggr]  
\end{eqnarray*}
\begin{eqnarray*}
     \textit{ Mediator Conditionnal } && \\
     \textit{Unconfoundness   } \hookrightarrow && \\
      &=&  \mathbb{E}_{C} \left[  \sum_{m  \in \mathcal{M}} \mathbb{E}  \left[Y \mid C, M = m, a, k \right ] \right. \\
      &&  \quad \quad \quad \mathbb{P}(M(a^{*}, p) = m \mid C, p, a^{*}) \mid j \biggr]  \\
          \textit{ Mediator Constistency  } 
            \hookrightarrow && \\
          &=&  \mathbb{E}_{C}\left[  \sum_{m  \in \mathcal{M}} \mathbb{E}  \left[Y \mid C, M = m, a, k \right ] \right. \\
          && \quad \mathbb{P}(M = m \mid C, a^{*}, p) \mid j \biggr]  \\
\end{eqnarray*}

By the MAR assumptions, we can express the components as:
\begin{align*}
\mathbb{P}(M=m \mid  a^*,C, p)
    &= \mathbb{P}(M=m \mid  a^*,C, p,\, R = 0), \\
\mathbb{E}[Y \mid a, M, C, k]
    &= \mathbb{E}[Y \mid a, M, C, k,\, R = 0] \\
    &= \mathbb{E}[Y \mid a, M, C, k,\, R = 0,\, R^* = 0].
\end{align*}

\section{Proof IPW formulation}
\subsection{IPW 1}

As we have

\begin{eqnarray*}
\mathbb{E}\left (Y \mid c, m, a ,k, R^*=0\right )
&=&
\frac{
\mathbb{E}\left (Y \, I(m)\, I(a)\, I(k)\, I(R^*=0)\mid c \right )
}{
\mathbb{P}(m \mid c, a,k, R=0)\,
\mathbb{P}(a \mid k, c, R=0)\,
\mathbb{P}(R=0 \mid c,k )
}
\frac{1}{\mathbb{P}(k \mid c )}
\end{eqnarray*}

\begin{eqnarray*}
\mathbb{E} \left[Y(a,k, M(a^{*}, p)) \mid j\right]
&=&
\sum_{(c,m)\in \mathcal{C} \times \mathcal{M}}
\Biggl(
\frac{
\mathbb{E}\left [Y I(m, a,k, R^*=0) \mid c  \right]
}{
\mathbb{P}(m \mid c, a,k, R=0)\,
\mathbb{P}(a \mid k, c)
}
\\
&& \qquad \times
\frac{\mathbb{P}(m \mid c, a^{*}, p, R=0)\,
\mathbb{P}(c \mid j)}{
\mathbb{P}(k \mid c, R=0)\,
\mathbb{P}(R^*=0 \mid m, a,k, c)\,
\mathbb{P}(R=0 \mid c,k )
}
\\
&=&
\frac{1}{\mathbb{P}(j)}
\sum_{(c,m)\in \mathcal{C} \times \mathcal{M}}
\Biggl(
\frac{
\mathbb{E}\left [Y I(m, a,k, R^*=0) \mid c  \right]
}{
\mathbb{P}(a \mid k, c, R=0)
}
\\
&& \qquad \times
\frac{\mathbb{P}(m \mid c, a^{*}, p, R=0)}
{\mathbb{P}(m \mid c, a,k, R=0)}
\,
\frac{\mathbb{P}(j \mid c)}
{\mathbb{P}(k \mid c)\,\mathbb{P}(R=0 \mid k, C)}
\, \mathbb{P}(c)
\Biggr)
\\[0.6em]
&=&
\frac{1}{\mathbb{P}(j)}
\sum_{m \in \mathcal{M}}
\mathbb{E}\Biggl[
\frac{
\mathbb{E}\left [Y I(m, a,k, R^*=0)\right]
}{
\mathbb{P}(a \mid k, C)
}
 \times
\frac{\mathbb{P}(m \mid C, a^{*}, p, R=0)}
{\mathbb{P}(m \mid C, a,k, R=0)}
\\
&& \qquad \times
\frac{\mathbb{P}(j \mid C)}
{\mathbb{P}(k \mid C)\,
\mathbb{P}(R=0 \mid k, C, a)\,
\mathbb{P}(R^*=0 \mid m, a,k, C, R=0)}
\Biggr]
\end{eqnarray*}

We also have:
\begin{eqnarray*}
\mathbb{E}\left [Y I(m, a, k, R^{*}=0)\right]
&=&
\mathbb{P}(M=m \mid C)\,
\mathbb{E}\left [Y I(a, k, R^{*}=0) \mid m \right] \\
\mathbb{P}(M \mid C, a,k, R=0)
&=&
\frac{
\mathbb{P}(a \mid C,M,k, R=0)\,
\mathbb{P}(k \mid C,M, R=0)
}{
\mathbb{P}(a \mid C,k, R=0)\,
\mathbb{P}(k\mid C, R=0)
}
\,
\mathbb{P}(M \mid C, R=0)
\end{eqnarray*}
\begin{eqnarray*}
\theta_{j,k,p;a,a^{*}}(P)
&=&
\frac{1}{\mathbb{P}(j)}
\mathbb{E}\Biggl[
\frac{
Y \, I(a, k, R^*=0)
}{
\mathbb{P}(a\mid C,M,k, R=0)\,
\mathbb{P}(k\mid C,M, R=0)
}
\\
&& \qquad \times
\frac{
\mathbb{P}(a^{*}\mid C,M, p,R=0)\,
\mathbb{P}(p\mid C,M, R=0)\,
\mathbb{P}(j\mid C)
}{
\mathbb{P}(a^{*}\mid C,p)\,
\mathbb{P}(p\mid C)\,
\mathbb{P}(R^*=0 \mid M, a,k, C, R=0)\,
\mathbb{P}(R=0 \mid a^*,p, C)
}
\Biggr].
\end{eqnarray*}
\subsection{IPW 2}
\begin{eqnarray*}
 \mathbb{E} \left[Y(a,k, M(a^{*}, p)) \mid j\right]  &=&  \mathbb{E}_{C}\Biggl [ \mathbb{E}_{M} \biggl[ \mathbb{E}  \Bigl[Y \mid C, M, a, k, R^{*}=0 \Bigr] \mid C,a^{*}, p , R=0\biggr]  \mid j \Biggr] \\
 &=&  \frac{1}{\mathbb{P}(j)} \mathbb{E} \Biggl[ \mathbb{E}\Biggl [ I(j) \mathbb{E} \biggl[ \mathbb{E}  \Bigl[Y \mid C, M, a, k, R^{*}=0 \Bigr] \mid C,a^{*}, p , R=0 \biggr]  \mid C \Biggr] \Biggr] \\
 &=&  \frac{1}{\mathbb{P}(j)} \mathbb{E} \Biggl[ \mathbb{E} \biggl[ \mathbb{E}  \Bigl[Y \mid C, M, a, k,  R^{*}=0 \Bigr] \mid C,a^{*}, p, R=0  \biggr]   \mathbb{P}(j  \mid C ) \Biggr]  \\
 &=& \frac{1}{\mathbb{P}(j)} \mathbb{E} \Biggl[ \mathbb{E} \biggl[ \frac{ I(a^{*}, p, , R=0 ) \mathbb{P}( j \mid C ) }{\mathbb{P}(a^{*} \mid C, p) \mathbb{P}(p \mid C) }\mathbb{E}  \Bigl[Y \mid C, M, a, k, R^{*}=0  \Bigr] \mid C \biggr]    \Biggr] \\
 &=&\frac{1}{\mathbb{P}(j)} \biggl[ \frac{ I(a^{*}, p,  R=0 ) \,   \mathbb{E} \left[Y \mid C, M, a, k, R^{*}=0  \right ]  \,  \, \mathbb{P}(j  \mid C )}{\mathbb{P}(  a^{*} \mid C, p,  R=0 )  \mathbb{P}(  R=0\mid C,  p)\mathbb{P}( p \mid C) }  \biggr] \\
\end{eqnarray*}

\section{Efficient Influent function}\label{proof_theoreme1}
By chain rule, we have :
\begin{eqnarray*}
    \frac{d}{dt} ( \theta(P)) = &&
   \overbrace{\displaystyle\sum_{(c,m) \in \mathcal{C} \times \mathcal{M}}  \color{magenta} \frac{d}{dt} \Biggr (\mathbb{E} \left[Y \mid c, m, a, k, R^*=0 \right ]  \Biggl ) \color{black} \mathbb{P}(m \mid c, a^{*}, p, R=0) \mathbb{P}(c \mid j )}^{=A} \\  + &&\displaystyle\sum_{(c,m) \in \mathcal{C} \times \mathcal{M}} \mathbb{E} \left[Y \mid c, m, a, k , R^*=0\right ] \color{blue}\frac{d}{dt} \Biggr (  \mathbb{P}(m \mid c, a^{*}, p, R=0) \Biggl ) \color{black}\mathbb{P}(c \mid j ) \Biggl \} =B \\+  && \underbrace{ \displaystyle\sum_{(c,m) \in \mathcal{C} \times \mathcal{M}} \mathbb{E} \left[Y \mid c, m, a, k, R^*=0 \right ]  
  \mathbb{P}(m \mid c, a^{*}, p,  R=0)  \color{teal}  \frac{d}{dt} ( \mathbb{P}(c \mid j ) ) }_{=C}
\end{eqnarray*}

It is well known that: 
\begin{eqnarray*}
     \color{magenta} \frac{d}{dt} \biggr (\mathbb{E} \left[Y \mid c, m, a, k, R^*=0 \right ]  \biggl ) = \frac{I(a, c, m, k, R^*=0)}{\mathbb{P}(a, c, m, k, R^*=0)} \biggl( Y - \mathbb{E} \left[Y \mid c, m, a, k, R^*=0 \right ]   \biggr)
\end{eqnarray*}

\begin{eqnarray*}
    \color{blue} \frac{d}{dt} \biggr (  \mathbb{P}(m \mid c, a^{*}, p, R=0 ) \biggl ) &=& \color{blue}  \frac{d}{dt} \biggr ( \mathbb{E}(I(m) \mid c, a^{*}, p, R=0) \biggl )\\  &=& \color{blue}\frac{I( c, a^{*}, p, R=0)}{\mathbb{P}( c, a^{*}, p, R=0)} \biggr ( I(m)  -  \mathbb{E}(I(m) \mid c, a^{*}, p, R=0)\biggl) \\  &=& \color{blue} \frac{I( c, a^{*}, p, R=0)}{\mathbb{P}( c, a^{*}, p, R=0)}  \biggr ( I(m)  -  \mathbb{P}(m \mid c, a^{*}, p, R=0)\biggl)
\end{eqnarray*}

\begin{eqnarray*}
    \color{teal} \frac{d}{dt} ( \mathbb{P}(c \mid j ) ) &=& \color{teal}   \frac{d}{dt} ( \mathbb{E}(I(c) \mid j ) ) = \frac{I(j)}{\mathbb{P}(j)} \times \biggr (I(c) -\mathbb{E}(I(c) \mid j )     \biggr)\\&=& \color{teal} \frac{I(j)}{\mathbb{P}(j)}  \biggr (I(c) -\mathbb{P}(c \mid j )     \biggr)
\end{eqnarray*}

So,

\begin{eqnarray*}
    A &=& \displaystyle\sum_{(c,m) \in \mathcal{C} \times \mathcal{M}}  \color{magenta} \frac{I(a, c, m, k, R^*=0)}{\mathbb{P}(a, c, m, k, R^*=0)} \biggl( Y - \mathbb{E} \left[Y \mid c, m, a, k,  R^*=0 \right ]   \biggr) \color{black} \mathbb{P}(m \mid c, a^{*}, p) \mathbb{P}(c \mid j ) \\
    &=& \frac{I(a, k, R^*=0)  \mathbb{P}(M \mid C, a^{*}, p,  R=0) \mathbb{P}(C \mid j )}{\mathbb{P}(a, C, M, k,  R^*=0)}  \biggl( Y - \mathbb{E} \left[Y \mid C, M, a, k,  R^*=0 \right ]   \biggr) 
   \\
    &=&  \frac{1}{\mathbb{P}(j)} \frac{
Y \, I(a,k, R^*=0) \, \mathbb{P}(a^{*}\mid C,M,  R=0) \,
}{
\mathbb{P}(a\mid C,M,  R=0) \, \mathbb{P}(k\mid C,M,  R=0) \, \mathbb{P}(a^{*}\mid C,p,  R=0) \,  \mathbb{P}( R=0 \mid C, p) 
}     \\
&& \times \frac{\mathbb{P}(p\mid C,M,  R=0) \, \mathbb{P}(j\mid C)}{\mathbb{P}(p\mid C) \, \mathbb{P}(R^*=0 \mid M, a,k, C,  R=0)}\biggl( Y - \mathbb{E} \left[Y \mid C, M, a, k, R^{*}=0 \right ]   \biggr)
\end{eqnarray*}

\begin{eqnarray*}
    B&=& \displaystyle\sum_{(c,m) \in \mathcal{C} \times \mathcal{M}} \mathbb{E} \left[Y \mid c, m, a, k, R^*=0 \right ] \color{blue}\frac{I( c, a^{*}, p, R=0)}{\mathbb{P}( c, a^{*}, p,  R=0)}  \biggr (  I(m)  -  \mathbb{P}(m \mid c, a^{*}, p,  R=0)\biggl) \color{black}\mathbb{P}(c \mid j ) \\
    &=& \frac{I( a^{*}, p, R=0)\mathbb{P}(C \mid j ) }{\mathbb{P}( C, a^{*}, p,  R=0)}  \biggr (  \mathbb{E} \left[Y \mid C, M, a, k, R^*=0 \right ] \\
    && \qquad \qquad \qquad \qquad \qquad \qquad  -  \sum_{m \in \mathcal{M} } \mathbb{E} \left[Y \mid C, m, a, k,  R^*=0  \right ] \mathbb{P}(m \mid C, a^{*}, p,  R=0 )\biggl) \\
     &=& \frac{1}{\mathbb{P}(j)}    \frac{ I(a^{*}, p,  R=0 )}{\mathbb{P}(  a^{*} \mid C, p,   R=0)  \mathbb{P}(  R=0 \mid  C, p,  )}\frac{\mathbb{P}(j  \mid C ) }{\mathbb{P}( p \mid C) }  \biggl(\mathbb{E} \left[Y \mid C, M, a, k,  R^*=0   \right ] \\
     && \qquad \qquad \qquad  \qquad \qquad \qquad - \mathbb{E}_{M} \biggl[ \mathbb{E} \left[Y \mid C, M, a, k,  R^*=0   \right ] \mid C,a^{*},p,  R=0  \biggr] \biggr) \\
\end{eqnarray*}

\begin{eqnarray*}
    C &=& \displaystyle\sum_{(c,m) \in \mathcal{C} \times \mathcal{M}} \mathbb{E} \left[Y \mid c, m, a, k, R^{*}=0 \right ]  
  \mathbb{P}(m \mid c, a^{*}, p, R=0)  \color{teal} \frac{I(j)}{\mathbb{P}(j)}  \biggr (I(c) -\mathbb{P}(c \mid j )     \biggr) \\  
  &=&  \frac{1}{\mathbb{P}(j)} \biggl [I(j) \displaystyle\sum_{m \in \mathcal{M}} \mathbb{E} \left[Y \mid C, m, a, k ,  R^{*}=0 \right ]  
  \mathbb{P}(m \mid C, a^{*}, p, R=0)  \\ && -  I(j)\displaystyle\sum_{(c,m) \in \mathcal{C} \times \mathcal{M}} \mathbb{E} \left[Y \mid c, m, a, k,  R^{*}=0  \right ]  
  \mathbb{P}(m \mid c, a^{*}, p, R=0)\mathbb{P}(c \mid j ) \Biggl]   \\  
   &=&  \frac{I(j)}{\mathbb{P}(j)} \biggl (  \mathbb{E}_{M} \Bigl[\mathbb{E} \left[Y \mid C, M, a, k,  R^{*}=0  \right ] \mid   C, a^{*}, p , R=0 \Bigr]
  -  \theta(P) \Biggl)   \\  
\end{eqnarray*}

\section{Double Robustness}

If $\widehat{Q}_{k}^{a}$ and $\widehat{b}_{k,p}^{a,a^*}$ are correctly specified and noting the other nuisance parameters,\[
\widehat{Q}_{k}^{a}(C,M)
=
\mathbb{E}(Y \mid C,M, A=a,S=k)
+o_{\mathbb{P}}(1)
\] \[
\widehat{b}_{k,p}^{a,a^*}(C)
=
\mathbb{E}\!\left(
\mathbb{E}(Y \mid C,M, A=a,S=k)
\mid C,A=a^*,S=p
\right)
+o_{\mathbb{P}}(1)
\]
and all the other parameter $\hat{\eta}$ converge to $\tilde{\eta}$
\[
\frac{
\widehat{\rho}_{a^*,p}(M_i,C_i)\,
\widehat{\tau}_{p}(C_i,M_i)\,
\widehat{p}_{j}(C_i)
}{
\widehat{\rho}_{a,k}(M_i,C_i)\,
\widehat{\tau}_{k}(C_i,M_i)\,
\widehat{\pi}_{a^*, p}(C_i)\,
\widehat{p}_p(C_i)
}
=
\tilde{w}_1 + o_{\mathbb{P}}(1)
\]
\[
\frac{1}{\, \widehat{\pi}_{a^{*}, p}(C_i) }
\frac{\, \widehat{p}_{j}(C_i)}{\, \widehat{p}_p(C_i)}
=
\tilde{w}_2 + o_{\mathbb{P}}(1)
\]

We also have that $\frac{n_j}{n}$ converges to $\mathbb{P}(S=j)$ and $O_i \sim P_0$.

\begin{eqnarray*}
&& \frac{ I(A_i=a,S_i=k)
\, \widehat{\rho}_{a^*,p}(M_i,C_i)\,
\widehat{\tau}_{p}(C_i,M_i)\,
\widehat{p}_{j}(C_i)
}{
\widehat{\rho}_{a,k}(M_i,C_i)\,
\widehat{\tau}_{k}(C_i,M_i)\,
\widehat{\pi}_{a^*, p}(C_i)\,
\widehat{p}_p(C_i)
}
\biggl( Y_i - \widehat{Q}_{k}^{a}(C_i,M_i) \biggr) \\
&& + \frac{I(A_i=a^{*}, S_i=p)}{\, \widehat{\pi}_{a^{*}, p}(C_i) }
\frac{\, \widehat{p}_{j}(C_i)}{\, \widehat{p}_p(C_i)}
\biggl( \widehat{Q}_{k}^{a}(C_i,M_i) - \widehat{b}_{k,p}^{a,a^*}(C_i) \biggr ) \\
&& + I(S_i=j)\widehat{b}_{k,p}^{a,a^*}(C_i) \\
&=&
I(A_i=a,S_i=k)\tilde{w}_1\biggl( Y_i - \mathbb{E}(Y_i \mid C_i,M_i, A=a,S_i=k) \biggr) \\
&& + I(A_i=a^{*}, S=p)\tilde{w}_2
\biggl(
\mathbb{E}(Y_i \mid C_i,M_i, A=a,S=k) - b_{k,p}^{a,a^*}(C_i)
\biggr) \\
&& + I(S_i=j)\mathbb{E}\!\bigl(\mathbb{E}(Y \mid C,M, A=a,S=k) \mid C, A=a^{*},S=p\bigr)
+ o_{\mathbb{P}}(1)
\end{eqnarray*}

So by the law of large numbers,
$\widehat{\theta}^{(\mathrm{one-step})}_{a, a^{*}}
\xrightarrow{\mathbb{P}}$

\begin{eqnarray*}
&&\frac{1}{\mathbb{P}(S=j)} \mathbb{E}_{P_0} \Biggl[
I(A=a,S=k)\tilde{w}_1\biggl( Y - \mathbb{E}(Y \mid C,M, A=a,S=k) \biggr) \\
&& \quad + I(A=a^{*}, S=p)\tilde{w}_2
\Biggl(
\mathbb{E}(Y \mid C,M, A=a,S=k)
- \\ && \quad \quad \quad \mathbb{E}\!\bigl(\mathbb{E}(Y \mid C,M, A=a,S=k)\mid C,A=a^{*},S=p\bigr)
\Biggr) \\
&& \quad + I(S=j)\mathbb{E}\!\bigl(\mathbb{E}(Y \mid C,M, A=a,S=k)\mid C,A=a^{*},S=p\bigr)
\Biggr]
\end{eqnarray*}

and the rest follows as in the derivation, yielding $\widehat{\theta}^{(\mathrm{one-step})}_{a, a^{*}} \xrightarrow{\mathbb{P}} \theta_{a,a^*}.$
\section{Data Application}

\begin{figure}
    \centering
    \includegraphics[angle=90,width=0.6\linewidth]{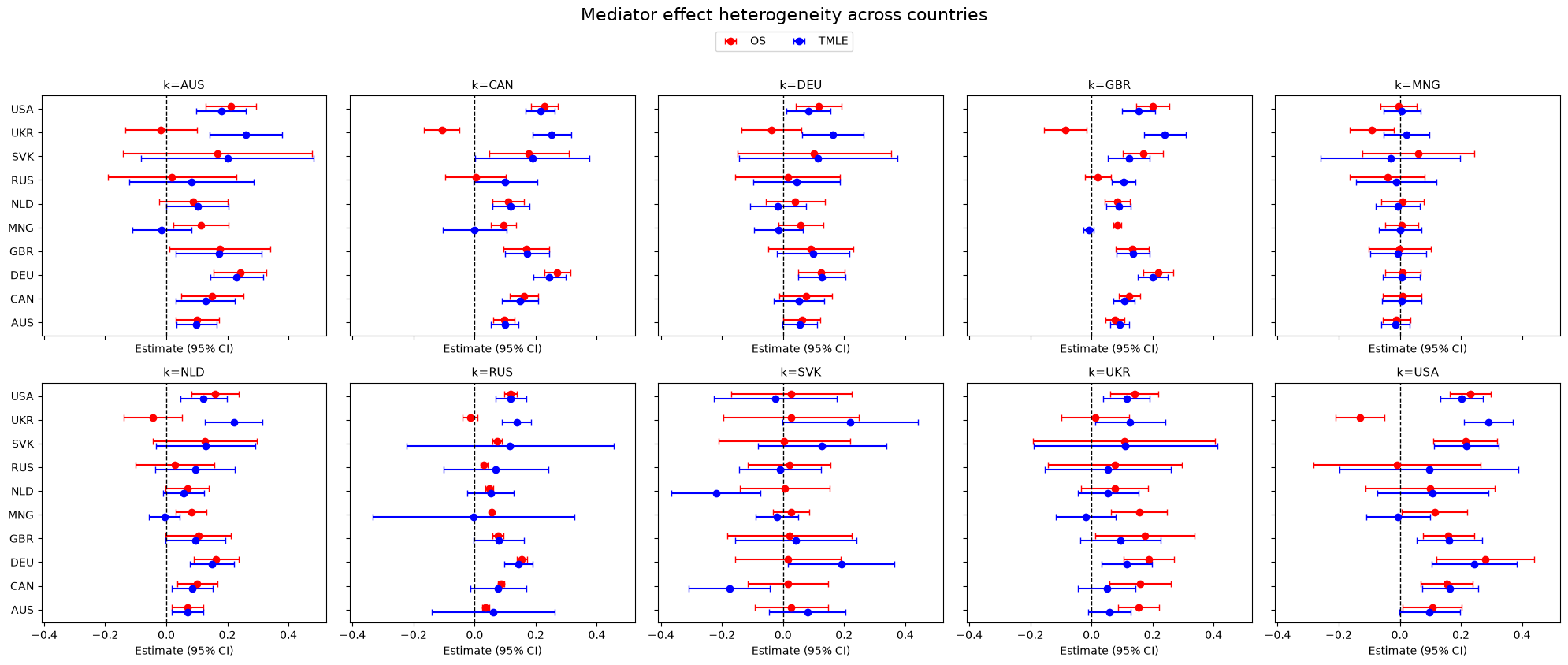}
    \caption{Treatment-Mediator heterogeneity across countries. Forest plot of the estimated effects for each country using TMLE and OS.}
    \label{fig:mediator_heterogeneity}
\end{figure}

\begin{figure}
    \centering
    \includegraphics[angle=90, width=0.6\linewidth]{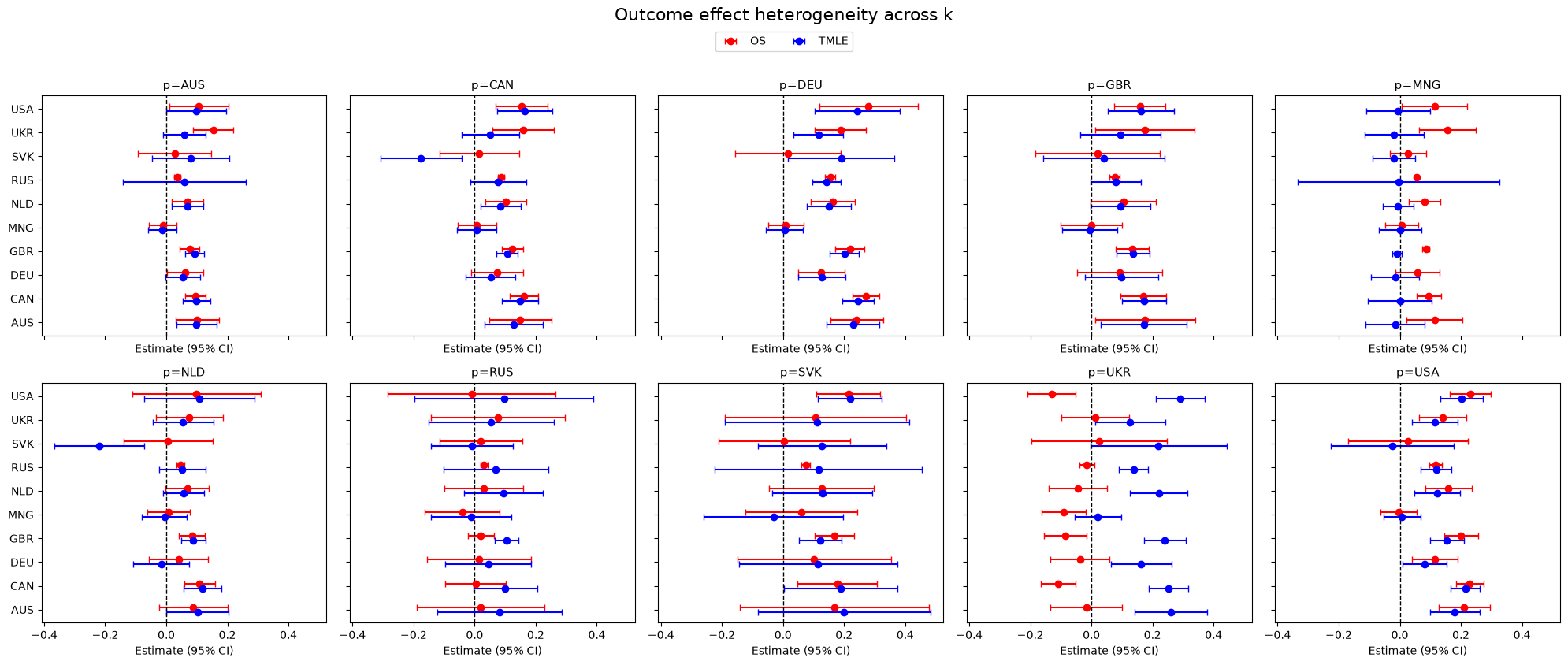}
 \caption{Treatment-Outcome heterogeneity across countries. Forest plot of the estimated effects for each country using TMLE and OS.}
    \label{fig:outcome_heterogeneity}
\end{figure}

\end{document}